\documentclass[12pt,a4paper]{article}

\usepackage{jheppub}

\usepackage{etoolbox}
    \makeatletter
    \patchcmd{\maketitle}{\@fpheader}{}{}{}
    \makeatother

\usepackage[utf8]{inputenc}
\usepackage{lmodern}
\usepackage{exscale}
\usepackage{graphicx}

\usepackage{amsmath}
\usepackage{amssymb}
\usepackage{bm}
\usepackage{mathtools}

\usepackage{mathrsfs}

\usepackage{relsize,exscale}
\usepackage{tensor}

\usepackage[low-sup]{subdepth}

\newcommand{\scr}{\scriptscriptstyle}
\newcommand{\bb}{\boldsymbol}

\usepackage{graphicx}

\usepackage{ulem}
\usepackage{cancel}

\usepackage[dvipsnames]{xcolor}
\hypersetup{
    colorlinks,
    linkcolor={black},
    citecolor={black},
    urlcolor={black}
}

\makeatletter
\newcommand{\dalembertian}{\mathop{\mathpalette\dalembertian@\relax}}
\newcommand{\dalembertian@}[2]{%
  \begingroup
  \sbox\z@{$\m@th#1\square$}%
  \dimen0=\fontdimen8
    \ifx#1\displaystyle\textfont\else
    \ifx#1\textstyle\textfont\else
    \ifx#1\scriptstyle\scriptfont\else
    \scriptscriptfont\fi\fi\fi3
  \makebox[\wd\z@]{%
    \hbox to \ht\z@{%
      \vrule width \dimen0
      \kern-\dimen0
      \vbox to \ht\z@{
        \hrule height \dimen0 width \ht\z@
        \vss
        \hrule height 2\dimen0
      }%
      \kern-2.5\dimen0
      \vrule width 2.5\dimen0
    }%
  }%
  \endgroup
}
\makeatother

\title{Appearances are deceptive: \\ Can graviton have a mass? }

\author[a]{Leihua Liu,}
\emailAdd{liuleihua8899@hotmail.com}
\author[b]{Tomislav Prokopec}
\emailAdd{t.prokopec@uu.nl}

\affiliation[a]{Department of Physics, College of Physics, Mechanical and Electrical Engineering, Jishou University, Jishou 416000, China
}
\affiliation[b]{Institute for Theoretical Physics, Spinoza Institute \& EMME$\Phi$,
	Utrecht University, 
	\\
	Princetonplein 5, 3584 CC Utrecht, The Netherlands}

\abstract{
We study the dynamics of linear gravitational perturbations on cosmological 
backgrounds of massive fermionic fields. We observe that, when gravitational and matter action
are expanded to quadratic order in gravitational perturbations on cosmological backgrounds,
the graviton appears to have an off-shell mass.
We derive a consistent set of two equations for the evolution of
linear classical and quantum gravitational perturbations on general cosmological backgrounds,
and demonstrate that the graviton mass disappears at the level of equations of motion (on-shell).
In the case we consider the expansion of the Universe is driven by 
the one-loop backreaction of fermions, and the dynamical gravitons evolve on the 
same background. 
These equations govern the evolution of linear gravitational perturbations on general 
cosmological matter backgrounds.
A concrete one-loop calculation is performed
for the simple case of massive Dirac fermions when the temperature of the cosmological fluid changes adiabatically when compared with the expansion rate of the Universe.
}

\notoc

\begin{document}

\maketitle

\titlepage

\section{Introduction}
\label{sec: Introduction}

The question of understanding the evolution of linearized gravitational
perturbations (and beyond) on general cosmological backgrounds
has gained importance after the onset 
of cosmic inflation~\cite{Starobinsky:1980te,Guth:1980zm}, 
and in particular 
after the discovery of gravitational waves~\cite{LIGOScientific:2016lio}. 
Namely, inflation
amplifies quantum matter density perturbations~\cite{Mukhanov:1981xt} and stretches them to vast cosmological scales,
such that they perturb homogeneous cosmic microwave background 
radiation~\cite{Planck:2019nip,Planck:2018jri}
and provide seeds to the Universe's large scale structure~\cite{DES:2022urg}. 

In this work we point out that insufficient attention has been given to 
proper understanding of the dynamics of linearlized gravitational perturbations on
quantum matter backgrounds.~\footnote{Textbooks~\cite{Mukhanov:2005sc,Weinberg:2008zzc} and 
reviews~\cite{Mukhanov:1990me} perform the analysis by assuming a fixed classical backgrounds, in which case there is no problem.} 
In particular, we point at the difficulties that arise when 
the na\^{i}ve approach is taken, namely when gravitational and matter actions are 
expanded to the second order in gravitational perturbations around general 
(spatially flat) cosmological backgrounds, whose expansion is driven by 
 quantized matter fields. We find that such a na\^{i}ve 
approach suggests that the dynamical graviton has a mass, which is inconsistent with
the common belief. In sections~\ref{On-shell analysis} 
and~\ref{Conservation laws come to the rescue} we slowly work towards the solution
of this disturbing observation, and show that the dynamical graviton, as expected,
possesses no mass. However, our analysis indicates that the constraint sector of the theory
couples to the non-vacuum part of the matter energy-momentum tensor. 
Rather than drawing any premature conclusions regarding what that signifies, 
we leave the precise interpretation of that coupling for a future investigation.

\section{Preliminaries}
\label{Preliminaries}

\subsection{Cosmological background}
\label{Cosmological background}

In this work we consider spatially homogeneous, isotropic and spatially flat cosmological spaces, whose geometry in general~$D$ dimensions
is characterised  by the conformally flat Friedman-Lema\^itre (FL) metric,
%
\begin{equation}
g^{\scr (0)}_{\mu\nu}\!=\! a^2(\eta) \, {\rm diag} (-1 , 1 , \dots , 1)
\,,
\label{metric: 0th order}
\end{equation}
where~$a$ is the scale factor that encodes the dynamics of the expansion,
and superscript $(0)$ denotes $0$-th order in the gravitational 
perturbations.

The rate of expansion is conveniently captured by the conformal Hubble rate,
\begin{equation}
\mathcal{H} = \frac{1}{a} \frac{{\rm d} a}{{\rm d} \eta} \, ,
\end{equation}
which is related to the physical Hubble rate as~$H\equiv \dot a/a \!=\! \mathcal{H}/a$, 
with $\dot a ={\rm d} a/{\rm d}t$ and ${\rm d}t = a{\rm d}\eta$.

From the definition of the Riemann tensor,
\begin{equation}
{R^\alpha}_{\mu\beta\nu} \!=\! \partial_\beta \Gamma^\alpha_{\mu\nu}
	\!-\! \partial_\nu \Gamma^\alpha_{\mu\beta}
	\!+\! \Gamma^\rho_{\mu\nu} \Gamma^\alpha_{\beta\rho}
	\!-\! \Gamma^\rho_{\mu\beta} \Gamma^{\alpha}_{\nu\rho} 
\,, 
\label{Riemann tensor}
\end{equation}
and the Christoffel connection,
\begin{equation}
\Gamma^\alpha_{\mu\nu} \!=\! \frac{1}{2} g^{\alpha\beta} ( \partial_\mu g_{\nu\beta}
	\!+\! \partial_\nu g_{\mu\beta} \!-\! \partial_\beta g_{\mu\nu} )
\,,
\label{Christoffel symbol}
\end{equation}
one can calculate
the curvature tensors in the homogeneous and isotropic FL spacetime~(\ref{metric: 0th order})
to be,
\begin{align}
R^{\scr (0)}_{\mu\nu\rho\sigma} ={}&a^2\big[
	2 \mathcal{H}^2 \eta_{\mu[\rho} \eta_{\sigma]\nu}
	-4(\mathcal{H}'\!-\!\mathcal{H}^2) 
     \bigl( a^2 \delta^0_{[\mu} g_{\nu] [ \sigma} \delta^0_{\rho]} \bigr)
\big]
	\, ,
\\
R^{\scr (0)}_{\mu\nu}\equiv g_{\scr (0)}^{\alpha\beta}
        R^{\scr (0)}_{\alpha\mu\beta\nu}  
   ={}&
	\big[(D\!-\!2)\mathcal{H}^2\!+\!\mathcal{H}' \big] \eta_{\mu\nu}
	- (D\!-\!2)\big(\mathcal{H}'\!-\!\mathcal{H}^2\big) \bigl(\delta_\mu^0 \delta_\nu^0 \bigr)
	\, ,
\\
R^{\scr (0)}\equiv g_{\scr (0)}^{\mu\nu} R^{\scr (0)}_{\mu\nu} ={}&a^{-2}
	(D\!-\!1)\big[(D\!-\!2)\mathcal{H}^2\!+\! 2\mathcal{H}'\big]
	\, .
\\
G_{\mu\nu}^{\scr (0)}\equiv  R^{\scr (0)}_{\mu\nu}
\!-\!\frac12g^{\scr (0)}_{\mu\nu} R^{\scr (0)}
 \!=\!{}&-\frac{D\!-\!2}{2}	
\big[(D\!-\!3)\mathcal{H}^2\!+\!2\mathcal{H}' \big] \eta_{\mu\nu}
	\!-\! (D\!-\!2)\big(\mathcal{H}'\!-\!\mathcal{H}^2\big) \bigl(\delta_\mu^0 \delta_\nu^0 \bigr)
	\, ,
\label{Einstein tensor: 0th order}
\end{align}
where the superscript ${(0)}$ stands for the zeroth order in gravitational perturbations.

The energy-momentum tensor of the matter field that drives the expansion
 has a perfect fluid form,
\begin{equation}
T^{\scr (0)}_{\mu\nu} = \big(\mathcal{P}^{\scr (0)}\!+\!\rho^{(0)}\big)u^{\scr (0)}_{\mu}u^{\scr (0)}_{\nu}
+  g^{\scr (0)}_{\mu\nu}\mathcal{P}^{(0)}
= a^2\big[ \big(\mathcal{P}^{\scr (0)}\!+\!\rho^{\scr (0)}\big)\delta^{0}_{\mu}
   \delta^{0}_{\nu}
+  \eta_{\mu\nu}\mathcal{P}^{\scr (0)}\big]
\label{perfect fluid}
\end{equation}
where $\mathcal{P}^{\scr (0)}$ and $\rho^{\scr (0)}$ are the energy density and pressure 
of the cosmological fluid and $u^{\scr (0)}_{\mu}=-a \delta_\mu^0$ is its velocity
(in the fluid rest frame), such that $g^{\mu\nu}_{\scr (0)} u^{\scr (0)}_{\mu}u^{\scr (0)}_{\nu} = -1$.

\subsection{Action}
\label{Action}

We are interested in the dynamics of the gravitational metric field $g_{\mu\nu}$, 
defined by the Hilbert-Einstein action in $D$ dimensions,
\begin{equation}
S_{\rm\scr HE}[g_{\mu\nu}] = \frac{1}{16\pi G}\int {\rm d}^Dx \sqrt{-g} 
 \big[R\!-\!(D\!-\!2)\Lambda_0\big]
\,,
\label{HE action}
\end{equation}
where, as usually, $R=g^{\mu\nu}R_{\mu\nu}$ is the Ricci curvature scalar, 
$R_{\mu\nu}$ the Ricci tensor, $g_{\mu\nu}$ the metric tensor,
$g^{\mu\nu}$ its inverse, $g={\rm det}[g_{\mu\nu}]$ and
$G$ and $\Lambda_0$ denote the (bare) Newton and cosmological constant,
respectively. We consider an expanding universe, whose expansion is
driven by a massive fermionic matter field $\psi$ ($\overline\psi=\psi^\dagger \gamma^0$), 
defined by the Dirac action, 
\begin{equation}
S_{\rm\scr D}[\psi,\overline{\psi},g_{\mu\nu}] = \int {\rm d}^Dx\sqrt{-g}\bigg[\frac{i}2 
   \big(\overline{\psi}e^\mu_{\,b}\gamma^b \nabla_\mu \psi
    -  (\nabla_\mu\overline{\psi})e^\mu_{\,b}\gamma^b  \psi\big)
    - \overline{\psi}M \psi
\bigg] 
\,,
\label{Dirac action}
\end{equation}
where $g = {\rm det}[g_{\mu\nu}]$, 
\begin{equation}
M  = m_R + i \gamma^5 m_I\,, \qquad (m_R, m_I \in \mathbb{R})
\,,
\label{fermionic mass}
\end{equation}
where $m_R$ and $m_I$ are fermionic mass and pseudo-mass parameters.
We allow separate scalar and pseudo-scalar masses, to more realistically model
the standard model, in which left- and right handed fermions have different dynamics.~\footnote{The left-handed fermions of the standard model are weakly charged and thus couple to the weak gauge bosons $W^\pm$ and $Z^0$, while the right-handed fermions do not.}
The fermionic action on curved spacetimes~(\ref{Dirac action}) 
is defined in terms of the tetrad vector field $e^\mu_{\,b}$, which connects 
space-time vectors with those on tangent space,
\begin{equation}
 \gamma^\mu(x) = e^\mu_{\,b}(x) \gamma^b
\,,
\label{Dirac gamma}
\end{equation}
where $\gamma^\mu(x)$ is the Dirac matrix on spacetime, which obeys
the usual anticommutation relation, 
\begin{equation}
\big\{ \gamma^\mu(x), \gamma^\nu(x)\big\} = - 2 g^{\mu\nu}(x)
\,,
\label{Dirac matrix: anticommutation relation}
\end{equation}
and $\gamma^b$ is the tangent space Dirac matrix, which obeys,  
the tangent-space anticommutation relation, 
\begin{equation}
\big\{ \gamma^b, \gamma^c\big\} = - 2 \eta^{bc}
\,,
\label{anticommutation relation: tangent space}
\end{equation}
in which there is no spacetime dependence. In this work, we use Greek letters $\mu,\nu, \cdots$ 
to denote spacetime indices on tensors, and Latin letters $b,c,\cdots$ to denote tangent space indices ($a$ is reserved for the scale factor). Next, $\nabla_\mu$ in~(\ref{Dirac action})
denotes the covariant derivative. It acts on the fermion field (and its conjugate) as
(see {\it e.g.}~\cite{Miao:2005am}),
\begin{equation}
\nabla_\mu \psi = (\partial_\mu - \Gamma_\mu)\psi
\,, \qquad \nabla_\mu \overline{\psi} = (\partial_\mu + \Gamma_\mu)\overline{\psi}
\,,
\label{covariant derivative on psi}
\end{equation}
where $\Gamma_\mu$ is the spin(or) connection given by,
\begin{equation}
\Gamma_\mu = -\frac{i}{2}\omega_{\mu cd}\sigma^{cd}
 \,,\qquad \omega_{\mu cd} = e^\nu_c\big(\partial_\mu e_{\nu d} 
  - \Gamma^\rho_{\mu\nu}e_{\rho d}\big)
\,,\quad
\sigma^{cd}= \frac{i}{4}\big[\gamma^c,\gamma^d\big]
\,.
\label{spinor connection}
\end{equation}
In what follows we expand the gravitonal and fermionic action to the second order 
in gravitational perturbations. 

\bigskip\noindent
{\bf Weyl transformations.} Since (massless) fermions couple conformally to gravity, it is convenient to employ Weyl transformations, 
\begin{equation}
g_{\mu\nu} = \Omega^2(x) \tilde g_{\mu\nu}
\,,\qquad  g =  \Omega^{2D}(x) \tilde g
\,,\qquad 
\psi(z) = \Omega^{-\frac{D-1}{2}}\chi(x)
\,.
\label{Weyl transformation}
\end{equation}
where 
\begin{eqnarray}
\tilde g_{\mu\nu} &\!=\!& \eta_{\mu\nu} + \kappa h_{\mu\nu}(x) 
\,, \qquad \tilde g = 1 + \frac{\kappa}{2} h
+ \frac{\kappa^2}{8} \big[h^2 -  2 h^{\mu\nu}h_{\mu\nu}\big] + {\cal O}(\kappa^3)
\,,
\nonumber\\
\tilde g^{\mu\nu}  &\!=\!& \eta^{\mu\nu} - \kappa h^{\mu\nu}(x)  
+ \kappa^2h^\mu_\rho (x)h^{\rho\nu}(x) + {\cal O}(\kappa^3)
\,,
\label{tilde gmn}
\end{eqnarray}
with $h^{\mu\nu}=\eta^{\mu\rho}\eta^{\nu\sigma}h_{\rho\sigma}$, 
$h = \eta^{\mu\nu}h_{\mu\nu}$ and 
\begin{equation}
\kappa^2 = 16\pi G
\label{kappa squared}
\end{equation}
is the loop-counting parameter of quantum gravity. 
Notice that for cosmology, $\Omega(x) \rightarrow a(\eta)$ reduces to the scale factor,
and that -- after Weyl rescaling -- the spatially flat background cosmological metric becomes (conformal) Minkowski metric $\eta_{\mu\nu}$. 

 From $g_{\mu\nu} = e_\mu^b \eta_{bc}e^c_\nu$, 
and $g^{\mu\nu} = e^\mu_b \eta^{bc}e_c^\nu$
it is a simple matter to show that 
tetrads transform as, 
\begin{equation}
e_\mu^b = \Omega(x) \tilde e_\mu^b
\,,\qquad 
e^\mu_b = \big[\Omega(x)\big]^{-1} \tilde e^\mu_b
\,,
\label{Weyl transformation: tetrads}
\end{equation}
where 
\begin{equation}
\tilde e_\mu^b = \delta_\mu^b + \frac{\kappa}{2} h_\mu^b 
-  \frac{\kappa^2}{8} h_\mu^\rho h_\rho^b + {\cal O}(\kappa^3)
\,,\quad 
\tilde e^\mu_b = \delta_\mu^b -\frac{\kappa }{2} h^\mu_b 
+  \frac{3\kappa^2}{8} h^\mu_\rho h^{\rho}_ b + {\cal O}(\kappa^3)
\,.\quad
\label{Weyl transformation: tetrads}
\end{equation}
Under Weyl transformations~(\ref{Weyl transformation})
 the Ricci tensor and scalar transform as
({\it cf.}~(\ref{Riemann tensor})), 
\begin{eqnarray}
R_{\mu\nu} &\!=\!& \tilde R_{\mu\nu} + \frac{1}{\Omega^2}
\left[2(D\!-\!2)(\partial_\mu \Omega)(\partial_\nu\Omega) 
 - (D\!-\!3)\tilde g^{\alpha\beta}(\partial_\alpha \Omega)(\partial_\beta\Omega)
  \tilde g_{\mu\nu}
  \right]
\nonumber\\
&&\hskip 1cm 
-\,\frac{1}{\Omega}\left[(D\!-\!2)(\widetilde\nabla_\mu \partial_\nu\Omega)
+ \tilde g_{\mu\nu}\widetilde{\dalembertian} \Omega
  \right]
\label{Weyl transformation: Ricci tensor}\\
R&\!=\!& \frac{1}{\Omega^2}\Big[
   \tilde R - \frac{1}{\Omega^2} (D\!-\!1) (D\!-\!4)\tilde g^{\mu\nu}(\partial_\mu\Omega)
       (\partial_\nu \Omega)
-2(D\!-\!1)\frac{1}{\Omega}\widetilde{\dalembertian} \Omega
                  \Big]
\,.
\label{Weyl transformation: Ricci scalar}
\end{eqnarray}

\medskip
\noindent
{\bf Gravitational action.} Upon inserting these into~(\ref{HE action}),
one obtains the gravitational action (up to $\kappa^2$),
\begin{equation}
S_{\rm\scr HE}[h_{\mu\nu}] = S^{\scr (0)}_{\rm\scr HE}[h_{\mu\nu}]  + S^{\scr (1)}_{\rm\scr HE}[h_{\mu\nu}] + S^{\scr (2)}_{\rm HE} [h_{\mu\nu}] 
 + {\cal O}(\kappa^3)
\,,
\label{expanded HE action}
\end{equation}
where 
\begin{eqnarray}
S^{\scr (0)}_{\rm\scr HE} &\!=\!&  \frac{1}{\kappa^2}\int\! {\rm d}^D x\, a^{D-2}
   \Big\{\!\!-(D\!-\!1)(D\!-\!2)\mathcal{H}^2 \!-\! (D\!-\!2)a^2\Lambda_0\Big\}
\,,
\label{HE: 0th order}\\
S^{\scr (1)}_{\rm\scr HE} &\!=\!&  \frac{1}{\kappa}\int\! {\rm d}^D x\, a^{D-2}
   \bigg\{h \frac{D\!-\!2}{2}\big[(D\!-\!3)\mathcal{H}^2 \!+\!2\mathcal{H}'\!-\!a^2\Lambda_0\big] 
\nonumber\\
&\!\!&\hskip 2.85cm
  \!+\, h^{\mu\nu}\delta^0_\mu\delta^0_\nu(D\!-\!2)\big(\mathcal{H}'\!-\!\mathcal{H}^2\big)\bigg\}
\,,\qquad\;
\label{HE: 1st order}\\
S^{\scr (2)}_{\rm\scr HE} &\!=\!& \int\! {\rm d}^D x\, a^{D-2}
   \bigg\{\frac12 \big(\partial^\rho h^{\mu\nu}\big)
                     \big(\partial_\mu h_{\rho\nu}\big)
\!-\!\frac12 \big(\partial^\mu h_{\mu\nu}\big)
                     \big(\partial^\nu h\big)
\!+\!\frac14 \big(\partial^\mu h\big)
                     \big(\partial_\mu h\big)
\nonumber\\
&\!\!&\hskip 2.45cm
\!-\,\frac14 \big(\partial^\rho h^{\mu\nu}\big)
                     \big(\partial_\rho h_{\mu\nu}\big)
\!-\!\frac{(D\!-\!2)}{2}\mathcal{H}\delta_\mu^0  h^{\mu\nu}
                     \big(\partial_\nu h\big)
\nonumber\\
&\!\!&\hskip 3.5cm
\!+\,\big(h^2\!-\!2 h^{\mu\nu}h_{\mu\nu}\big)\frac{D\!-\!2}{8}
                 \big[(D\!-\!3)\mathcal{H}^2\!+\!2\mathcal{H}'\!-\!a^2\Lambda_0\big]
\nonumber\\
&\!\!&\hskip 4.5cm
\!+\,\big(hh^{\mu\nu}\!-\!2 h^{\mu\rho} h^\nu_{\rho}\big)\delta^0_\mu\delta^0_\nu
                      \frac{D\!-\!2}{2}
                 \big(\mathcal{H}'\!-\!\mathcal{H}^2\big)\!
\bigg\}
\,.
\qquad\;
\label{HE: 2nd order}
\end{eqnarray}
When $D\rightarrow 4$ this action reduces to the one derived in Rick Vinke's thesis~\cite{RickVinke:2020}, and agrees with~\cite{Tsamis:1992xa}, 
 up to no-derivative terms quadratic in $h_{\mu\nu}$, which were omitted 
in~\cite{Tsamis:1992xa} on the account 
of the 0th order Einstein's equation in de Sitter, $(D-1)\mathcal{H}^2 - a^2 \Lambda_0 =0$.
Notice that the quadratic terms in the last two lines 
of~(\ref{HE: 2nd order}) drop out in the de Sitter limit, in which $\mathcal{H}'\!-\!\mathcal{H}^2=0$, as they should. Furthermore, in the de Sitter limit also $S^{(1)}_{\rm\scr HE}\rightarrow 0$.

\medskip
\noindent
{\bf Fermion action.} Massless fermions couple conformally to gravity in 
arbitrary dimension, which means that upon a Weyl transformation, 
the Dirac action~(\ref{Dirac action}) reduces to, 
\begin{equation}
S_{\rm\scr D}[\chi,\overline{\chi},h_{\mu\nu}] = \int {\rm d}^Dx\sqrt{-\tilde g}
 \bigg[\frac{i}2 
   \big(\overline{\chi}\tilde e^\mu_{\,b}\gamma^b \widetilde\nabla_\mu \chi
    -  (\widetilde{\nabla}_\mu\overline{\chi})\tilde e^\mu_{\,b}\gamma^b  \chi\big)
    - a\overline{\chi}M \chi
\bigg] 
\,,
\label{Dirac action: Weyl}
\end{equation}
where $\widetilde\nabla_\mu  \chi$ denotes the covariant derivative with respect to
 $\tilde g_{\mu\nu}$.
It is now relatively simple to expand this action in powers of $h_{\mu\nu}$. The result is 
%
\begin{equation}
S_{\rm\scr D}[\chi,\overline{\chi},h_{\mu\nu}] = S^{\scr (0)}_{\rm\scr D}[\chi,\overline{\chi}] + S^{\scr (1)}_{\rm\scr D}[\chi,\overline{\chi},h_{\mu\nu}]+ S^{\scr (2)}_{\rm\scr  D} [\chi,\overline{\chi},h_{\mu\nu}]
 + {\cal O}(\kappa^3)
\,,
\label{expanded Dirac action}
\end{equation}
where
\begin{eqnarray}
S^{\scr (0)}_{\rm\scr D} &\!=\!& \int {\rm d}^Dx
 \bigg[\frac{i}2 
   \big(\overline{\chi}\gamma^\mu \partial_\mu \chi
    -  (\partial_\mu\overline{\chi})\gamma^\mu  \chi\big)
    - a\overline{\chi}M \chi
\bigg] \,,
\label{Dirac action: 0th order}\\
S^{\scr (1)}_{\rm\scr D} &\!=\!& \frac{\kappa}{2}\int {\rm d}^Dx
 \Big[h \mathcal{L} - h^{\mu\nu} \mathcal{K}_{\mu\nu}
\Big] 
\,,
\label{Dirac action: 1st order}\\
S^{\scr (2)}_{\rm\scr D} &\!=\!& \frac{\kappa^2}{8}\int {\rm d}^Dx
 \bigg[\big(h^2\!-\!2h^{\mu\nu}h_{\mu\nu}\big) \mathcal{L}
   \!+\!\big(3h^\mu_\rho h^{\rho\nu}\!-\!2hh^{\mu\nu}\big)\mathcal{K}_{\mu\nu}
\nonumber\\
&\!\!&\hskip 2.1cm
\!+\,\frac12 h^\sigma_\mu\partial_\nu h_{\sigma\rho}\overline{\chi}\big\{\gamma^\nu,\sigma^{\mu\rho}\big\}\chi
\bigg] 
\,,\qquad
\label{Dirac action: 2nd order}
\end{eqnarray}
where for notational convenience we introduced the kinetic Dirac operator and Lagrangian,
\begin{eqnarray}
\mathcal{K}_{\mu\nu} &\!\equiv\!&
 \frac{i}2 
   \big(\overline{\chi}\gamma_\nu \partial_\mu \chi
    -  (\partial_\mu\overline{\chi}) \gamma_\nu \chi\big)
\,,\qquad\\
\mathcal{L} &\!\equiv\!&\eta^{\mu\nu}\mathcal{K}_{\mu\nu}
  - a \overline{\chi}M\chi
\,,
\label{definition of Kmn and L}
\end{eqnarray}
where we assumed that $\{\gamma^b,\sigma^{cd}\}$ is antisymmetric in all three indices 
(which is rigorously true in $D=4$, where 
$\{\gamma^b,\sigma^{cd}\} = -\epsilon^{ebcd}\gamma_e\gamma^5$).
In~(\ref{Dirac action: 0th order})--(\ref{definition of Kmn and L})  and from now on we use condensed notation for the Dirac matrices on tangent space, 
$\gamma^\mu \equiv \delta^\mu_b \gamma^b$.
The action~(\ref{Dirac action: 0th order})--(\ref{Dirac action: 2nd order}) agrees with~\cite{RickVinke:2020}, provided one replaces,
$\big\{\gamma^\nu,\sigma^{\mu\rho}\big\}\rightarrow 
- \epsilon^{\alpha\nu\mu\rho}\gamma_\alpha\gamma^5$, which is legitimate in $D=4$.

\medskip
\noindent
{\bf The 2PI fermionic action.}
In this work we are primarily interested in fermionic fluids that drive the expansion of the Universe, and for that purpose the dynamics of fermions is better captured 
by the action defined in terms of the fermionic two-point functions, whose definitions
are recalled in what follows.

The positive and negative frequency fermionic
Wightman two-point functions are defined as,
\begin{equation}
iS^{-+}(x;x') = {\rm Tr}\left[\hat\rho_{\scr in}\hat\psi(x)\hat{\overline{\psi}}(x')\right]
\,,\qquad
iS^{+-}(x;x') = -{\rm Tr}\left[\hat\rho_{\scr in}\hat{\overline{\psi}}(x')\hat\psi(x)\right]
\,,\quad
\label{2 point Wightman function}
\end{equation}
where $\hat\rho_{\scr in}$ denotes the (initial) density operator (in Heisenberg picture)
and we adapted the Keldysh notation (the minus sign in the negative frequency Wightman function is due to the anticommuting nature of 
fermions). The Feynman (time ordered) propagator is then,
\begin{equation}
iS^{++}(x;x') = \Theta(t\!-\!t') iS^{-+}(x;x')+ \Theta(t'\!-\!t) iS^{+-}(x;x')
\,,\quad
\label{Feynman propagator}
\end{equation}
where  $\Theta(x)$ denotes the Heaviside theta function, which acts as projector and divides spacetime into the future and the past sections. The Dyson (anti-time ordered) propagator is obtained by exchanging the Wightman functions in~(\ref{2 point Wightman function}).
The Wightman functions obey homogeneous Dirac equations, while 
the propagator satisfies,~\footnote{The propagator equation of 
motion~(\ref{Feynman propagator equation}) can be derived from the canonical anticommutation relation,
\begin{equation}
\big\{\hat\psi(t,\vec x),\hat\Pi_\psi(t,\vec x^{\,\prime})\big\}
= i\hbar \delta^{D-1} (\vec x\!-\!\vec x^{\,\prime})
\,,
\end{equation}
where $\hat\Pi_\psi(t,\vec x')$ is the canonical momentum 
operator defined by, $\Pi_\psi(x)=\delta S_{\rm \scr D}/\delta\partial_0\psi(x) 
= \sqrt{-g} \psi^\dagger ie^0_b(x)\gamma^0\gamma^b$, and keeping in mind 
that the Wightman functions obey homogeneous equations of motion.
}
\begin{equation}
\sqrt{-g}\big(ie^\mu_b \gamma^b\nabla_\mu \!-\! M\big)iS^{++}(x;x') 
   = i \hbar\delta^D(x\!-\!x')
\,.\quad
\label{Feynman propagator equation}
\end{equation}
The propagator (as well as the Wightman functions) obeys a second equation on the $x'$ leg, which can be replaced by
the symmetry requirement on the propagator, $iS^{++}(x;x') =iS^{++}(x';x)$.
The Dyson propagator obeys  the same equation, but with an opposite sign 
in front of the delta function. For the purposes of this work, it is more convenient to 
use the Weyl rescaled fields $\chi$ and $\overline{\chi}$~(\ref{Weyl transformation}), 
and define the two-point functions 
in terms of these,
\begin{equation}
i\tilde S^{-+}(x;x') = {\rm Tr}\left[\hat\rho_{\scr in}\hat\chi(x)\hat{\overline{\chi}}(x')\right]
\,,\qquad
i\tilde S^{+-}(x;x') = -{\rm Tr}\left[\hat\rho_{\scr in}\hat{\overline{\chi}}(x')\hat\chi(x)\right]
\,,\quad
\label{2 point Wightman function}
\end{equation}
in terms of which we can define the corresponding Feynman and Dyson propagators,
\begin{eqnarray}
i\tilde S^{++}(x;x') &\!=\!&
   \Theta(t\!-\!t') i\tilde S^{-+}(x;x')+ \Theta(t'\!-\!t) i\tilde S^{+-}(x;x')
\label{Feynman propagator: rescaled}\\
i\tilde S^{--}(x;x') &\!=\!& \Theta(t\!-\!t') i\tilde S^{+-}(x;x')+ \Theta(t'\!-\!t) i\tilde S^{-+}(x;x')
\,.\quad
\label{Dyson propagator: rescaled}
\end{eqnarray}
We are now ready write the fermionic two-particle irreducible (2PI) action, 
which can be organized similarly as the classical action~(\ref{Dirac action: 0th order})--(\ref{Dirac action: 2nd order}),
\begin{equation}
\Gamma_{\rm\scr D}[iS^{\tt bc},h^{\tt b}_{\mu\nu}] = S^{(0)}_{\rm\scr D}[iS^{\tt bc}] 
+ S^{\scr (1)}_{\rm\scr D}[iS^{\tt bc},h^{\tt b}_{\mu\nu}]
+ S^{\scr (2)}_{\rm\scr  D} [iS^{\tt bc},h^{\tt b}_{\mu\nu}]
+ \Gamma^{\scr (1)}_{\rm\scr D}[iS^{\tt bc}] 
 + {\cal O}(\kappa^3)
\,,
\label{expanded Dirac action: 2PI}
\end{equation}
where 
\begin{eqnarray}
S^{\scr (0)}_{\rm\scr D} &\!=\!& \sum_{{\tt b,c}=\pm}\!\!{\tt b}\! \int {\rm d}^Dx{\rm d}^Dx'
\bigg[\!-\frac{i}2 \gamma^\mu \big(\partial_\mu \!-\! \partial_\mu^\prime\big)
    \!+\! aM
\bigg]\delta^D(x\!-\!x')\delta^{\tt bc}i S^{{\tt bc}}(x';x) 
\,,\qquad
\label{Dirac action: 0th order: 2PI}\\
&\!\equiv\!& \sum_{{\tt b,c}=\pm}\!\!{\tt b}\,\delta^{\tt bc}\! \int {\rm d}^Dx
\big[\mathcal{L}^{{\tt bc}}(x;x')\big]_{x'\rightarrow x} 
\nonumber\\
S^{\scr (1)}_{\rm\scr D} &\!=\!& \frac{\kappa}{2}
 \sum_{{\tt b,c}=\pm}\!\!{\tt b}\,\delta^{\tt bc}\!\int {\rm d}^Dx
 \Big\{h^{\tt b} \big[\mathcal{L}^{{\tt bc}}(x;x')\big]_{x'\rightarrow x}
   - h_{\tt b}^{\mu\nu}\big[\mathcal{K}^{{\tt bc}}_{\mu\nu}(x;x')\big]_{x'\rightarrow x}
\Big\}
\,,\qquad
\label{Dirac action: 1st order: 2PI}\\
S^{\scr (2)}_{\rm\scr D} &\!=\!& \frac{\kappa^2}{8} 
\sum_{{\tt b,c}=\pm}\!\!{\tt b}\,\delta^{\tt bc}\!\int {\rm d}^Dx
 \bigg\{\big(h_{\tt b}^2\!-\!2h_{\tt b}^{\mu\nu}h^{\tt b}_{\mu\nu}\big)
\big[\mathcal{L}^{{\tt bc}}(x;x')\big]_{x'\rightarrow x}
\nonumber\\
&\!\!&\hskip 3.7cm
   \!+\,\big(3(h^{\tt b})^{\mu}_\rho h_{\tt b}^{\rho\nu}\!-\!2h^{\tt b}h_{\tt b}^{\mu\nu}\big)
 \big[\mathcal{K}^{{\tt bc}}_{\mu\nu}(x;x')\big]_{x'\rightarrow x}
\nonumber\\
&\!\!&\hskip 3.7cm
\!-\,\frac12 (h^{\tt b})^{\,\sigma}_\mu\big(\partial_\nu h^{\tt b}_{\sigma\rho}\big)\big\{\gamma^\nu,\sigma^{\mu\rho}\big\}\big[i S^{{\tt bc}}(x';x)\big]_{x'\rightarrow x}
\!\bigg\}
\,,\qquad
\label{Dirac action: 2nd order: 2PI}
\end{eqnarray}
and $\Gamma^{\scr (1)}_{\rm\scr D}$ denotes the one-loop contribution,
\begin{equation}
\Gamma_{\rm\scr D}^{\scr (1)} = \sum_{{\tt b,c}=\pm}
 \int {\rm d}^Dx\, {\rm d}^Dx^\prime \Big\{ i\hbar {\rm Tr}\left[\ln\big(S^{{\tt bc}}(x';x)\big)\right]\!\Big\}
\,.
\label{Dirac action: 1 loop: 2PI}
\end{equation}
We have used typewritter Latin letters for Keldysh indices,
and we introduced a shorthand notation, 
\begin{eqnarray}
\mathcal{K}_{\mu\nu}^{{\tt bc}}(x;x')&\equiv&
-\frac{i}2 \gamma_\nu \big(\partial_\mu \!-\! \partial_\mu^\prime\big)i S^{{\tt bc}}(x';x) 
\,,\qquad\\
\mathcal{L}^{{\tt bc}}(x;x') &\equiv&
  \eta^{\mu\nu}\mathcal{K}^{{\tt bc}}_{\mu\nu}(x;x')
  + a Mi S^{{\tt bc}}(x';x) 
\,,
\label{definition of Kmn and L: 2PI}
\end{eqnarray}
which are the nonlocal generalizations of~(\ref{definition of Kmn and L})
suitable for the 2PI formalism. Notice that a trace over the spinor indices is implied
in~(\ref{Dirac action: 0th order: 2PI})--(\ref{definition of Kmn and L: 2PI}). 

The gravitational action~(\ref{expanded HE action})--(\ref{HE: 2nd order})
can be in analogous fashion adapted to the Schwinger-Keldysh notation. As here we are not interested in the graviton loops, 
this is a trivial procedure, as it entails promoting the field 
$h_{\mu\nu}\rightarrow h^{{\tt b}}_{\mu\nu}$ and adding a factor 
${\tt b}$ in front of the action and summing over ${\tt b}$, so we will not explicitly write it.
Instead, we will focus on the structure of the gravitational and fermionic actions. 
Notice first that the first order action~(\ref{HE: 1st order}) can be recast as, 
\begin{eqnarray}
S^{\scr (1)}_{\rm\scr HE} &\!=\!&  \frac{1}{\kappa}\sum_{{\tt b}=\pm}{\tt b}\int\! {\rm d}^D x\, a^{D-2}
   \Big\{\!\!-h_{\tt b}^{\mu\nu}\Big(G_{\mu\nu}^{\scr (0)}
                              \!+\! \frac{D\!-\!2}{2}a^2\eta_{\mu\nu}\Lambda_0\Big) \Big\}
\,,\qquad\;
\label{HE: 1st order: simple}
\,.
\qquad\;
\end{eqnarray}
where $G_{\mu\nu}^{\scr (0)}$ is the background 
Einstein tensor~(\ref{Einstein tensor: 0th order}). The fermionic first order action
\begin{eqnarray}
S^{\scr (1)}_{\rm\scr D} &\!=\!& \frac{\kappa}{2}
 \sum_{{\tt b}=\pm}{\tt b}\!\int {\rm d}^Dx
 \Big\{h_{\tt b}^{\mu\nu}\Big( \eta_{\mu\nu} \big[\mathcal{L}^{{\tt bb}}(x;x')\big]_{x'\rightarrow x}
   - \big[\mathcal{K}^{{\tt bb}}_{\mu\nu}(x;x')\Big)\big]_{x'\rightarrow x}
\Big\}
\nonumber\\
&\!\equiv\!&\hskip 0cm
\, \frac{\kappa}{2}
 \sum_{{\tt b}=\pm}{\tt b}\!\int {\rm d}^Dx \Big\{h_{\tt b}^{\mu\nu} {T^{\scr (0){\tt b}}}_{\!\mu\nu}\Big\}
\,,\qquad
\label{Dirac action: 1st order: 2PI b}
\end{eqnarray}
where ${T^{\scr (0){\tt b}}}_{\!\mu\nu}$ is the background energy-momentum tensor for fermions. 
Varying the first order action with respect to $h^{\mu\nu}_{\tt b}$ gives the background 
equation of motion, 
\begin{equation}
G_{\mu\nu}^{\scr (0)}\!+\! \frac{D\!-\!2}{2}a^2\Lambda_0=\frac{\kappa^2}{2}  T_{\mu\nu}^{\scr (0)} 
=\frac{\kappa^2}{2} a^{2-D} \Big( \eta_{\mu\nu}\mathcal{L}(x;x) - \mathcal{K}_{\mu\nu}(x;x)\Big)
\,,\qquad
\label{EE: background}
\,.
\quad
\end{equation}
where we used that on-shell ${T_{\tt b}^{(0)}}_{\!\mu\nu}\rightarrow {T^{(0)}}_{\!\mu\nu}$,
is independent of ${\tt b}$, and so are $\mathcal{L}(x;x)$ and $\mathcal{K}_{\mu\nu}(x;x)$
evaluated at spacetime coincidence. (This is of course true provided $\mathcal{L}(x;x)\rightarrow \mathcal{L}^{\scr (0)}(x;x) $ and 
$\mathcal{K}_{\mu\nu}(x;x)\rightarrow \mathcal{K}^{\scr (0)}(x;x)$ are solved to
the leading order in $\kappa$.)
The same is true of the graviton field: on-shell 
$h^{\mu\nu}_{\tt b}\rightarrow h^{\mu\nu}$.
 This also means that,
when the Einstein equation is enforced in the first order action, it vanishes.  In other words, 
when Einstein equation is enforced 
the linear graviton action does not induce any tadpole source for the gravitons,
and that the solution to the Einstein equation is the stationary point of the gravitational action.  

Let us now consider the second order action. The gravitational part of interest are the zero-derivative terms
in the last two lines of~(\ref{HE: 2nd order}), which in the Keldysh rendering are,
\begin{eqnarray}
S^{(2)}_{\rm\scr HE} &\!\supset\!& \sum_{{\tt b}=\pm}{\tt b}\!\!\int\! {\rm d}^D x\, a^{D-2}
   \bigg\{
\big(h_{\tt b}^2\!-\!2 h_{\tt b}^{\mu\nu}h^{\tt b}_{\mu\nu}\big)\frac{D\!-\!2}{8}
                 \big[(D\!-\!3)\mathcal{H}^2\!+\!2\mathcal{H}'\!-\!a^2\Lambda_0\big]
\nonumber\\
&\!\!&\hskip 4.5cm
+\,\big(h^{\tt b}h_{\tt b}^{\mu\nu}\!-\!2 h_{\tt b}^{\mu\rho} (h_{\tt b})^\nu_{\rho}\big)\delta^0_\mu\delta^0_\nu
                      \frac{D\!-\!2}{2}
                 \big(\mathcal{H}'\!-\!\mathcal{H}^2\big)\!
\bigg\}
\,.
\qquad\;
\label{HE: 2nd order: Keldysh}
\end{eqnarray}
The term in the first line contains the $\eta_{\mu\nu}$ part of $G_{\mu\nu}^{\scr (0)}$ minus the cosmological constant constribution,
while the term in the second line contains the $\delta_\mu^0\delta_\nu^0 $ 
part of $G_{\mu\nu}^{\scr (0)}$
in~(\ref{Einstein tensor: 0th order}). 
Consider now the fermionic second order action~(\ref{Dirac action: 2nd order: 2PI}).
Eqs.~(\ref{EE: background}) and~(\ref{Einstein tensor: 0th order}) imply that, at zeroth order
in $\kappa$, 
$-\mathcal{K}_{\mu\nu}(x;x)$ can be split as, 
\begin{equation}
-\mathcal{K}^{\scr (0)}_{\mu\nu}(x;x) = \mathcal{Q}^{\scr (0)} \eta_{\mu\nu} + \delta_\mu^0\delta_\nu^0  \mathcal{S}^{\scr (0)}
\,,\quad
\label{Kmn: split}
\end{equation}
such that
\begin{equation}
\mathcal{P}^{\scr (0)} = a^{-D} \big(\mathcal{L}^{\scr (0)} + \mathcal{Q}^{\scr (0)}\big)
\,,\quad
\mathcal{P}^{\scr (0)}+\rho^{\scr (0)} = a^{-D}\mathcal{S}^{\scr (0)}
\,,
\label{pressure and rho vs L and K}
\end{equation}
denote the pressure and 
entropy density $\mathcal{P}^{\scr (0)} + \rho^{\scr (0)}$ 
(up to a factor of temperature)~\cite{BarrosoMancha:2020fay} 
(recall that the entropy density of a thermal fluid is 
${\it s}^{\scr (0)} = (\mathcal{P}^{\scr (0)} + \rho^{\scr (0)})/T$).
The form~(\ref{Kmn: split}) is also dictated by the symmetries of the FL spacetime. Namely, 
if one evaluates the coincident fermionic two-point function in a state obeying the cosmological symmetries, 
the split in~(\ref{Kmn: split}) is the most general one consistent with translational and rotational symmetries of cosmological spaces. Upon inserting Eq.~(\ref{Kmn: split}) 
into~(\ref{Dirac action: 2nd order: 2PI})
 one obtains, 
\begin{eqnarray}
S^{\scr (2)}_{\rm\scr D} &\supset& \frac{\kappa^2}{8} 
\sum_{{\tt b,c}=\pm}\!\!{\tt b}\,\delta^{\tt bc}\!\int {\rm d}^Dx
 \bigg\{\big(h_{\tt b}^2\!-\!2h_{\tt b}^{\mu\nu}h^{\tt b}_{\mu\nu}\big)
\mathcal{L}^{{\tt bc}}(x;x)
   \!+\!\big(2h^2_{\tt b}\!-\!3h_{\tt b}^{\mu\nu} h^{\tt b}_{\mu\nu}\big)
\mathcal{Q}^{\scr (0)}_{\tt b}
\nonumber\\
&\!\!&\hskip 3.7cm
   \!+\,\big(2h^{\tt b}h_{\tt b}^{\mu\nu}\!-\!3(h_{\tt b})_{\rho}^{\mu} (h_{\tt b})^{\rho\nu}\big)
 \big(\delta_\mu^0\delta_\nu^0  \mathcal{S}^{\scr (0)}_{\tt b}\big)
\!\bigg\}
\,,\qquad\;
\label{Dirac action: 2nd order: 2PI b}
\end{eqnarray}
where we restored Keldysh indices on $\mathcal{Q}^{\scr (0)}$ and $\mathcal{S}^{\scr (0)}$
and we dropped the last line, as it contains derivatives acting on $h_{\mu\nu}$
(the tensorial structure of~(\ref{Kmn: split}) should be maintained off shell  
for coincident $\mathcal{K}^{\scr (0)}_{\mu\nu}(x;x)$). 
Now comparing~(\ref{HE: 2nd order: Keldysh}) with~(\ref{Dirac action: 2nd order: 2PI b})
and in light of~(\ref{pressure and rho vs L and K}), one concludes 
that the zero-derivative terms do not cancel when the background equation of 
motion~(\ref{EE: background}) is used.
This then means that off-shell the graviton appears to have a mass. Indeed, summing the two
contributions  yields, 
\begin{eqnarray}
S^{(2)}_{\rm\scr HE} &\!\supset\!& \sum_{{\tt b}=\pm}{\tt b}\!\!\int\! {\rm d}^D x\, a^{D-2}
   \bigg\{
\frac{h_{\tt b}^2\!-\!2 h_{\tt b}^{\mu\nu}h^{\tt b}_{\mu\nu}}{4}\bigg[\frac{D\!-\!2}{2}
                 \big[(D\!-\!3)\mathcal{H}^2\!+\!2\mathcal{H}'\!-\!a^2\Lambda_0\big]
\nonumber\\
&\!\!&\hskip 4.5cm
+ \frac{\kappa^2}{2}\Big( \mathcal{L}^{{\tt bc}}(x;x)\!+\!\mathcal{Q}^{\scr (0)}\Big)\bigg]+\frac{\kappa^2}{8} \big(h^2_{\tt b}\!-\!h_{\tt b}^{\mu\nu} h^{\tt b}_{\mu\nu}\big)
\mathcal{Q}^{\scr (0)}
\nonumber\\
&\!\!&\hskip -0.6cm
+\,\frac{h^{\tt b}h_{\tt b}^{\mu\nu}\!-\!2 h_{\tt b}^{\mu\rho} (h_{\tt b})^\nu_{\rho}}{2}\delta^0_\mu\delta^0_\nu
                     \bigg[ (D\!-\!2)
                 \big(\mathcal{H}'\!-\!\mathcal{H}^2\big)
\!+\!\frac{\kappa^2}{2}  \mathcal{S}^{\scr (0)}\bigg]
\!+\!\frac{\kappa^2}{8}(h_{\tt b})_{\rho}^{\mu} (h_{\tt b})^{\rho\nu}\big)
 \big(\delta_\mu^0\delta_\nu^0  \mathcal{S}^{\scr (0)}\big)
\!
\bigg\}
\nonumber\\
&\!\!&\hskip -0.cm
\longrightarrow \sum_{{\tt b}=\pm}{\tt b}\!\!\int\! {\rm d}^D x\, a^{D-2}
   \bigg\{
\frac{\kappa^2}{8} \big(h^2_{\tt b}\!-\!h_{\tt b}^{\mu\nu} h^{\tt b}_{\mu\nu}\big)
\mathcal{Q}^{\scr (0)}
\!+\!\frac{\kappa^2}{8}(h_{\tt b})_{\rho}^{\mu} (h_{\tt b})^{\rho\nu}\big)
 \big(\delta_\mu^0\delta_\nu^0  \mathcal{S}^{\scr (0)}\big)
\!
\bigg\}
.
\qquad\;\;
\label{HE plus Dirac : 2nd order: Keldysh}
\end{eqnarray}
The fact that these terms do not drop out means that the graviton have an off-shell mass,
which was first noted in~\cite{RickVinke:2020}, supervised by one of the authors.
In the following section we ask the question whether this mass 
survives on-shell, which would imply a physical mass. 
The expectation is that somehow the would-be graviton mass terms 
in~(\ref{HE plus Dirac : 2nd order: Keldysh}) cancel on-shell.

\section{On-shell analysis}
\label{On-shell analysis}

In this section we analyse the equations of motion for fermions and 
gravitons~\footnote{At the linear level, the equations of motion for classical gravitational 
waves and potentials is identical to its quantum counterpart.} with the goal to resolve 
the question of the graviton mass. As the reader will see, the results are somewhat surprising.

\subsection{Equations of motion}
\label{Equations of motion}

Varying the fermionic action~(\ref{Dirac action: 0th order: 2PI})--(\ref{Dirac action: 1 loop: 2PI}) 
with respect to $iS^{{\tt cb}}(x^\prime;x)$, and multiplying from the right by it,
gives the equation of motion for the fermionic two-point functions,
\begin{eqnarray}
&&\hskip -0.3cm
\Big\{\!\big(i\gamma^\mu\partial_\mu \!-\! aM\big)
\!+\!\frac{\kappa}{2}\Big[h\big(i\gamma^\mu\partial_\mu \!-\! aM\big)
\!-\!h^{\nu}_\mu i\gamma^\mu\partial_\nu
\!+\!\frac12\big(\partial_\mu h\!-\!\partial_\nu h^{\nu}_\mu\big)i\gamma^\mu\Big]
         \Big\} i\tilde S^{{\tt bc}}(x;x') 
\nonumber\\
&&\hskip 9.9cm
= i\hbar (\sigma^3)^{{\tt bc}}\delta^D(x\!-\!x')
\,,\qquad
\label{Dirac equation for rescaled field}
\end{eqnarray}
where $\sigma^3 = {\rm diag}(1,-1)$ is the Pauli matrix.
By multiplying from the left, one 
obtains that the fermions also satisfy the equation on the second leg,
\begin{eqnarray}
&&\hskip -0.3cm
 i\tilde S^{{\tt bc}}(x;x') \Big\{\!\big(\!-\!i\gamma^\mu\overleftarrow{\partial}^\prime_\mu \!-\! a^\prime M\big)
\!+\!\frac{\kappa}{2}\Big[\big(\!-\!\overleftarrow{\partial}^\prime_\mu i\gamma^\mu \!-\! a^\prime M\big)h(x')
\nonumber\\
&&\hskip 2.9cm
\!+\, i\gamma^\mu\overleftarrow{\partial}^\prime_\nu h^{\nu}_\mu(x')
\!-\!\frac12\big(\partial^\prime_\mu h\!-\!\partial^\prime_\nu h^{\nu}_\mu(x')\big)i\gamma^\mu\Big]
         \!\Big\} = i\hbar (\sigma^3)^{{\tt bc}}\delta^D(x\!-\!x')
\,,\qquad\;
\label{Dirac equation for rescaled field: leg x'}
\end{eqnarray}
The second equation is obtained by applying hermitian conjugation on
the first one~(\ref{Dirac equation for rescaled field}).
%
%

 Assuming we know the solution
of the leading order equation,
\begin{equation}
\big(i\gamma^\mu\partial_\mu \!-\! aM\big) i\tilde S_{{\tt bc}}^{\scr (0)}(x;x') 
= i\hbar (\sigma^3)^{{\tt bc}}\delta^D(x\!-\!x')
\,,\qquad
\label{Dirac equation zeroth order}
\end{equation}
one can write the general solution of~(\ref{Dirac equation for rescaled field})
as,~\footnote{The na\^\i ve solution of~(\ref{Dirac equation for rescaled field})
is 
\begin{eqnarray}
 i\tilde S^{{\tt bc}}(x;x') &\!\equiv\!&
 i\tilde S^{\scr (0)}_{{\tt bc}}(x;x')\!+\!\kappa\, i\tilde S^{\scr (1)}_{{\tt bc}}(x;x')
  = \left(1\!-\!\frac{\kappa}{2}h(x^\prime)\right) i\tilde S_{{\tt bc}}^{\scr (0)}(x;x')
\nonumber\\
&\!+\!&\hskip 0.cm
\frac{\kappa}{2}\!\int\! {\rm d}^Dy\, G_{\scr R}(x;y)i\gamma^\mu
\Big[h^{\nu}_\mu(y) \partial^y_\nu
\!-\!\frac12\big(\partial_\mu h(y)\!-\!\partial_\nu h^{\nu}_\mu(y)\big)\Big]
 i\tilde S_{{\tt bc}}^{\scr (0)}(y;x')
\,,\qquad
\label{naive solution Dirac equation}
\end{eqnarray}
where  $G_{\scr R}(x;y)$ is the retarded Green's function obeying, 
\begin{equation}
\big(i\gamma^\mu\partial_\mu \!-\! aM\big) G_{\scr R}(x;x^\prime) = \delta^D(x\!-\!x^\prime) 
\,.\quad
\label{retarded Green function eom}
\end{equation}
The retarded Green's function can be expressed in terms of the two-point functions as, 
\begin{equation}
G_{\scr R}(x;x') 
=i\tilde S^{\tt ++}(x;x') - i\tilde S^{\tt -+}(x;x') 
 = \Theta(\eta\!-\!\eta^\prime)\big[i\tilde S^{\tt +-}(x;x') \!-\! i\tilde S^{\tt -+}(x;x')\big]
\,.\quad
\label{fermionic retarded Green's function in Keldysh}
\end{equation}
Solution~(\ref{naive solution Dirac equation})
is not entirely correct however, as it does not satisfy 
the Dirac equation~(\ref{Dirac equation for rescaled field: leg x'}) on the $x^\prime$ leg.
This is so because the Feynman and Dyson propagators are not causal Green's functions 
and receive contributions both when $x'$ is in the future and in the past of $x$.
}
\begin{eqnarray}
 i\tilde S^{{\tt bc}}(x;x') &\!\equiv\!&
 i\tilde S^{\scr (0)}_{{\tt bc}}(x;x')\!+\!\kappa\, i\tilde S^{\scr (1)}_{{\tt bc}}(x;x') 
=i\tilde S_{{\tt bc}}^{\scr (0)}
\nonumber\\
&&\hskip -1.8cm
 - \,\frac{\kappa}{2i\hbar}\sum_{{\tt de}}
\!\int\! {\rm d}^Dy\, i\tilde S_{{\tt bd}}^{\scr (0)}(x;y)\Big[ h(y)
\big(i\gamma^\mu\partial^y_\mu \!-\! aM\big)\!+\!\frac12i\gamma^\mu\big(\partial^y_\mu h(y)\big)\Big](\sigma^3)^{{\tt de}}
          i\tilde S_{{\tt ec}}^{\scr (0)}(y;x') 
\nonumber\\
&\!\!&\hskip -0.9cm
\!+\,
\frac{\kappa}{2i\hbar}\!\sum_{{\tt de}}
\!\int\! {\rm d}^Dy\, i\tilde S_{{\tt bd}}^{\scr (0)}(x;y)
   i\gamma^\mu
\Big[h^{\nu}_\mu(y) \partial^y_\nu
\!+\!\frac12\big(\partial_\nu h^{\nu}_\mu(y)\big)\Big]
(\sigma^3)^{{\tt de}} i\tilde S_{{\tt ec}}^{\scr (0)}(y;x')
\,,\quad\;
\label{Dirac equation: solution}
\end{eqnarray}
where $ i\tilde S^{\scr (0)}_{{\tt bc}}(x;x')$ and $ i\tilde S^{\scr (1)}_{{\tt bc}}(x;x')$
denote the 0th and 1st order in $\kappa$ fermionic two-point functions, respectively. 
By integrating by parts the $y$-derivatives in~(\ref{Dirac equation: solution}) 
acting on the fermionic two-point functions, the solution can be equivalently written as, 
\begin{eqnarray}
 i\tilde S^{{\tt bc}}(x;x') &\!\equiv\!&
 i\tilde S^{\scr (0)}_{{\tt bc}}(x;x')\!+\!\kappa\, i\tilde S^{\scr (1)}_{{\tt bc}}(x;x') 
=i\tilde S_{{\tt bc}}^{\scr (0)}
\nonumber\\
&&\hskip -1.9cm
 - \,\frac{\kappa}{2i\hbar}\sum_{{\tt de}}
\!\int\! {\rm d}^Dy\, i\tilde S_{{\tt bd}}^{\scr (0)}(x;y)\Big[
\big(\!-\!i\gamma^\mu\overleftarrow{\partial}^y_\mu \!-\! aM\big) h(y)\!-\!\frac12i\gamma^\mu\big(\partial^y_\mu h(y)\big)\Big](\sigma^3)^{{\tt de}}
          i\tilde S_{{\tt ec}}^{\scr (0)}(y;x') 
\nonumber\\
&\!\!&\hskip -1.3cm
\!+\,
\frac{\kappa}{2i\hbar}\!\sum_{{\tt de}}
\!\int\! {\rm d}^Dy\, i\tilde S_{{\tt bd}}^{\scr (0)}(x;y)
   i\gamma^\mu
\Big[\!-\!\overleftarrow{\partial}^y_\nu h^{\nu}_\mu(y) 
\!-\!\frac12\big(\partial_\nu h^{\nu}_\mu(y)\big)\Big]
(\sigma^3)^{{\tt de}} i\tilde S_{{\tt ec}}^{\scr (0)}(y;x')
\,,\qquad\;\;
\label{Dirac equation: solution: leg x'}
\end{eqnarray}
which solves the equation of motion on the $x'$ leg~(\ref{Dirac equation for rescaled field: leg x'}),
showing that both equations of motion are satisfied by the solution~(\ref{Dirac equation: solution}).
The na\^{\i}ve solution~(\ref{naive solution Dirac equation}) does not have that property.


On cosmological spaces, the general spinor structure of the 0th order two-point functions is,
\begin{equation} 
i\tilde S^{\scr(0)}_{{\tt bc}}(x;x')  = \big(i\gamma^\mu\partial_\mu \!+\! aM^\dagger\big)
\Big[P^\varphi_{+}i\tilde \Delta_{\scr (+)}^{{\tt bc}}(x;x')
\!+\!P^\varphi_{-}i\tilde \Delta_{\scr (-)}^{{\tt bc}}(x;x') \Big]
\,,\quad
\label{Ansatz: fermionic 2pt fn: 0th order}
\end{equation}
where $P^\varphi_\pm\equiv \big(1\mp e^{i\varphi\gamma^5}\gamma^0\big)/2$
($\tan(\varphi)=m_R/m_I$) are 
the (rotated) projectors on particle and antiparticle states 
(positive and negative frequency shells)~\cite{Prokopec:2022yon}, 
and the scalar two-point functions 
$i\tilde \Delta_{\scr (\pm)}^{{\tt bc}}(x;x')$ obey,
\begin{equation}
\big(\partial^2 - a^2m^2 \mp i a\mathcal{H}m\big)i\tilde \Delta_{\scr (\pm)}^{{\tt bc}}(x;x')
 =  i\hbar (\sigma^3)^{{\tt bc}}\delta^D(x\!-\!x')
\,,\quad
\label{equations for particles and antiparticles}
\end{equation}
where $m^2 = m_R^2 + m_I^2$ is the fermionic mass squared.
These equations cannot be solved for general cosmological backgrounds $a=a(\eta)$. Nevertheless,
they can be solved in some simple cases, for example in radiation era (ultra-relativistic fluids), in which the equation of state parameter is $w=\mathcal{P}^{\scr (0)}/\rho^{\scr (0)} = 1/3$, and in which
$a=A_0 \eta$ ($A_0 ={\rm const.}$), as well as in general backgrounds when $m=0$.~\footnote{The propagators are known on some fixed cosmological backgrounds, such as de Sitter~\cite{Candelas:1975du}, 
\cite{Prokopec:2022yon}; and on spaces with constant acceleration,
in which $\epsilon = -\dot H/H^2 = 1-\mathcal{H}'/\mathcal{H}^2 ={\rm const.}$, 
in which the mass parameter $m^2$ is proportional to the Ricci curvature 
scalar~\cite{Koksma:2009tc}.}
But more on that later.

\medskip

Next, varying~(\ref{HE: 0th order})--(\ref{HE: 2nd order}) 
and~(\ref{Dirac action: 0th order: 2PI})--(\ref{Dirac action: 2nd order: 2PI})  with respect to the gravitational field yields the equation of motion 
for the metric perturbations,
\begin{eqnarray}
G_{\mu\nu}^{\scr (0)}
\!+\!\kappa\mathcal{L}_{\mu\nu}^{\;\;\;\;\rho\sigma}h_{\rho\sigma}
\!+\!\frac{D\!-\!2}{2}a^2\Lambda_0 \eta_{\mu\nu} 
\!+\!\kappa \frac{D\!-\!2}{2}a^2\Lambda_0h_{\mu\nu}
&\!=\!& \frac{\kappa^2}{2}T_{\mu\nu}
\nonumber\\
&&\hskip -10.2cm
 \equiv \frac{\kappa^2}{2} a^{2-D}
\Big\{\eta_{\mu\nu} \mathcal{L}(x;x)\!-\! \mathcal{K}_{\mu\nu}(x;x)
\label{Einstein equation for perturbations}\\
&\!\!& \hskip -10.2cm
+\,\kappa\Big[
                        h_{\mu\nu}\mathcal{L} \!-\! \frac{1}{2}h_{(\mu}^\beta \mathcal{K}_{\nu)\beta} 
\!-\! \frac12\eta_{\mu\nu} h^{\alpha\beta} \mathcal{K}_{\alpha\beta}
\!+\!\frac14 \epsilon^{\tau\sigma\beta\delta}\big(\partial_\sigma h_{\beta(\mu}\big)
  \eta_{\nu)\delta}\gamma_\tau \gamma^5 iS^{{\tt bb}}(x;x)
                \Big]
\Big\}
\,,
\nonumber
\end{eqnarray}
where we dropped Keldysh indices, as the dependence on 
them drops out from one-point functions 
and coincident two-point functions, representing expectation values of composite operators 
(this is because two-point functions
 do not contain imaginary parts when evaluated at coincidence).
The operator $\mathcal{L}^{\mu\nu\rho\sigma}$ in~(\ref{Einstein equation for perturbations})
is known as the Lichnerowicz operator,
which on cosmological spaces acts as~\cite{Park:2015kua},
\begin{eqnarray}
\mathcal{L}_{\mu\nu}^{\;\;\;\;\rho\sigma}h_{\rho\sigma} 
&\!=\!& \partial_\rho \partial_{(\mu}h^\rho_{\nu)}
\!-\!\frac12\partial^2h_{\mu\nu}
\!-\!\frac12\partial_{\mu}\partial_{\nu}h
\!-\!\frac12\eta_{\mu\nu}\partial^\rho \partial^\sigma h_{\rho\sigma}
\!+\!\frac12 \eta_{\mu\nu}\partial^2h
\nonumber\\
&\!-\!&\hskip 0.cm
(D\!-\!2)\mathcal{H}\Big(\partial_{(\mu}h_{\nu)0}\!-\!\frac12\partial_0h_{\mu\nu}\Big)
\!+\!(D\!-\!2)\mathcal{H}\Big(\partial^\rho h_{\rho 0}\!-\!\frac12\partial_0h\Big)\eta_{\mu\nu}
\nonumber\\
&\!-\!&\hskip 0.cm\frac{D\!-\!2}{2}\big((D\!-\!3)\mathcal{H}^2 \!+\!2\mathcal{H}^\prime\big)
\big(h_{\mu\nu}\!+\!h_{00}\eta_{\mu\nu}\big)
\,.\qquad
\label{Lichnerowicz operator}
\end{eqnarray}
Equations~(\ref{Einstein equation for perturbations})--(\ref{Lichnerowicz operator}) 
can be used to establish whether there is an on-shell graviton mass. 
First, note that at the leading order in $\kappa$ (not counting the classical coupling
 $\kappa^2/2 = 8\pi G$), the left- 
and right-hand-sides nicely combine into the  0th order Einstein equation~(\ref{EE: background}),
and therefore -- to the order $\kappa^0$ -- it is all fine.
Next to consider are the terms linear in $\kappa$. For clarity, we write them again explicitly,
\begin{eqnarray}
\mathcal{L}^{\mu\nu\rho\sigma}h_{\rho\sigma}
\!-\!\frac{D\!-\!2}{4}a^2\Lambda_0 \big(h\eta_{\mu\nu}\!-\!2h_{\mu\nu}\big) 
&\!=\!& \frac{\kappa^2}{2}T^{\scr (1)}_{\mu\nu}
\nonumber\\
&&\hskip -5.2cm
 \equiv \frac{\kappa^2}{2} a^{2-D}
\Big\{\eta_{\mu\nu} \mathcal{L}^{\scr (1)}(x;x)\!-\! \mathcal{K}^{\scr (1)}_{\mu\nu}(x;x)
\label{Einstein equation for perturbations: 1st order}\\
&\!\!& \hskip -5.2cm
+\,\Big[
                        h_{\mu\nu}\mathcal{L}^{\scr (0)} \!-\! \frac{1}{2}h_{(\mu}^\beta 
\mathcal{K}^{\scr (0)}_{\nu)\beta} 
\!-\! \frac12\eta_{\mu\nu} h^{\alpha\beta} \mathcal{K}^{\scr (0)}_{\alpha\beta}
\!+\!\frac14 \epsilon^{\tau\sigma\beta\delta}\big(\partial_\sigma h_{\beta(\mu}\big)
  \eta_{\nu)\delta}\gamma_\tau \gamma^5 iS^{\scr (0)}_{{\tt bb}}(x;x)
                \Big]
\Big\}
\,,\quad\,
\nonumber
\end{eqnarray}
where $i\tilde S^{\scr (1)}_{{\tt bb}}(x;x')$ is defined in~(\ref{Dirac equation: solution}).
Now collecting the terms in~(\ref{Einstein equation for perturbations: 1st order}), 
(\ref{Lichnerowicz operator})
with no derivatives acting on $h_{\mu\nu}$ results in the following 
left- and right-hand sides,
\begin{eqnarray}
(LHS)_{\tt \scr 0\, der.}&\!=\!&-\frac{D\!-\!2}{2}\big((D\!-\!3)\mathcal{H}^2 \!+\!2\mathcal{H}^\prime\big)
\big(h_{\mu\nu}\!+\!h_{00}\eta_{\mu\nu}\big)
\nonumber\\
&&\hskip 0.2cm
\!+\,\frac{D\!-\!2}{2}a^2\Lambda_0 \Big(h_{\mu\nu}\!-\!\frac12h\eta_{\mu\nu}\Big) 
\label{EE: 1st order no derivatives LHS}\\
(RHS)_{\tt \scr 0\,der.} &\!=\!&\frac{\kappa^2}{2} a^{2-D}
\Big\{\eta_{\mu\nu} \mathcal{L}^{\scr (1){\tt bb}}(x;x)
 \!-\! \mathcal{K}^{\scr (1){\tt bb}}_{\mu\nu}(x;x)
\nonumber\\
&\!\!& \hskip 0.cm
+\,\Big[h_{\mu\nu}\mathcal{L}^{\scr (0)} \!+\! \frac{1}{2}
\big(\mathcal{Q}^{\scr (0)}h_{\mu\nu}
 \!-\!h_{0(\mu}\delta_{\nu)}^0\mathcal{S}^{\scr (0)}\big)
\!+\! \frac12\eta_{\mu\nu} \big(\mathcal{Q}^{\scr (0)}h\!+\!h_{00}\mathcal{S}^{\scr (0)}\big)
\Big\}
\,,\qquad
\label{EE: 1st order no derivatives RHS}
\end{eqnarray}
where 
\begin{eqnarray}
 \mathcal{K}^{{\tt bb}\scr (1)}_{\mu\nu}(x;x) &\!=\!& 
-\frac{i}{2}\big[\gamma_{(\mu}\left(\partial_{\nu)}\!-\!\partial_{\nu)}^\prime\right)
   i\tilde S^{\scr (1)}_{{\tt bb}}(x;x')\big]_{x'\rightarrow x}
\,,
\label{Kmn 1st order}\\
 \mathcal{L}_{{\tt bb}}^{\scr (1)}(x;x) 
&\!=\!& \eta^{\mu\nu}\mathcal{K}^{{\tt bb}\scr (1)}_{\mu\nu}(x;x)
    + aM iS^{\scr (1)}_{{\tt bb}}(x;x)
\,,\qquad
\label{L 1st order BB}
\end{eqnarray}
and we used~(\ref{Kmn: split}) and
\begin{eqnarray}
 i\tilde S^{{\tt(1)}}_{{\tt bc}}(x;x') 
&&
\nonumber\\
&&\hskip -1.8cm
= - \frac{1}{2i\hbar}\sum_{{\tt de}}
\!\int\! {\rm d}^Dy\, i\tilde S_{{\tt bd}}^{\scr (0)}(x;y)\Big[ h(y)
\big(i\gamma^\mu\partial^y_\mu \!-\! aM\big)\!+\!\frac12i\gamma^\mu\big(\partial^y_\mu h(y)\big)\Big](\sigma^3)^{{\tt de}}
          i\tilde S_{{\tt ec}}^{\scr (0)}(y;x') 
\nonumber\\
&\!\!&\hskip -0.9cm
\!+\,
\frac{1}{2i\hbar}\!\sum_{{\tt de}}
\!\int\! {\rm d}^Dy\, i\tilde S_{{\tt bd}}^{\scr (0)}(x;y)
   i\gamma^\mu
\Big[h^{\nu}_\mu(y) \partial^y_\nu
\!+\!\frac12\big(\partial_\nu h^{\nu}_\mu(y)\big)\Big]
(\sigma^3)^{{\tt de}} i\tilde S_{{\tt ec}}^{\scr (0)}(y;x')
\,.\quad\;
\label{Dirac equation: solution: 1st order}
\end{eqnarray}
Let us first have a look at the dynamical gravitons, that is the terms containing $h_{\mu\nu}$ in 
Eqs.~(\ref{EE: 1st order no derivatives LHS})--(\ref{EE: 1st order no derivatives RHS}).
The left-hand-side~(\ref{EE: 1st order no derivatives LHS}) is just 
$h_{\mu\nu}$ times the Lorentz covariant part of the Einstein equation 
({\it i.e.} the part multiplying $\eta_{\mu\nu}$), 
the right-hand-side~(\ref{EE: 1st order no derivatives RHS}) is,
\begin{equation}
(RHS)_{\scr 0\, der.}\supset
\frac{\kappa^2}{2} a^{2-D} h_{\mu\nu}\Big[\mathcal{L}^{\scr (0)} \!+\! \frac{1}{2}
                                \mathcal{Q}^{\scr (0)}
        \Big]
\,,\qquad
\label{RHS 0 der hmn}
\end{equation}
plus the nonlocal terms containing $\mathcal{K}^{\scr(1){\tt bb}} _{\mu\nu}$ and 
$\mathcal{L}_{{\tt bb}}^{\scr (1)}$. The 0th order Einstein equation~(\ref{EE: background}), (\ref{Kmn: split}) 
contains a factor $1$ in the last term, instead of $1/2$ in~(\ref{RHS 0 der hmn}), so unless
one can extract another factor $1/2$ from the nonlocal terms, one would have to conclude that 
the graviton is massive on-shell. The situation in the scalar sector 
(the terms containing $h_{00}$ and $h$) 
is even more complex: not even the left-hand-side agrees with the leading order 
Einstein equation. Since we do not see a simple way to localize $h_{\mu\nu}(y)$ 
in the nonlocal terms, it still appears that the dynamical graviton is massive on-shell.

To summarize, we have found that the non-local form of the graviton equation 
prevents us from removing the mass term from both the dynamical
and scalar part of the graviton equation of 
motion~(\ref{Einstein equation for perturbations: 1st order}).


\section{Conservation laws come to the rescue}
\label{Conservation laws come to the rescue}

Apart from the equations of motion, Einstein's gravity provides us with three conservation laws,
the contracted Bianchi identity, metric compatibility and the energy-momentum conservation,
\begin{equation}
\nabla^\mu G_{\mu\nu} = 0\,,\qquad 
\nabla^\mu g_{\mu\nu} = 0\,,\qquad 
\nabla^\mu T_{\mu\nu} = 0
\,.\qquad
\label{conservation laws}
\end{equation} 
These act as consistency conditions and could provide extra information that may be
useful. So let us check that. 

\medskip
{\bf Energy-momentum tensor.}
Upon writing $T_{\mu\nu} = a^{2-D}\widetilde T_{\mu\nu}$
and acting the covariant derivative on the right-hand-side 
of~(\ref{Einstein equation for perturbations}) yields (to the first order in $\kappa$),  
\begin{eqnarray}
&&\hskip -0.4cm
a^{-D}\Big\{\big[\partial^\mu \widetilde T_{\mu\nu} 
\!-\! \mathcal{H}\delta_\nu^0 \eta^{\mu\nu}\widetilde T_{\mu\nu}\big] 
\!-\!\kappa\Big[\partial_\mu\big(h^{\mu\rho}\widetilde T_{\rho\nu}\big)
   \!+\!\frac12(\partial_\nu h^{\rho\mu})\widetilde  T_{\rho\mu}
    \!-\!\frac12(\partial^\mu h)\widetilde  T_{\mu\nu}\Big]
\nonumber\\    
&&\hskip 9.8cm      
\!+\, \kappa\mathcal{H}\delta_\nu^0h^{\mu\rho}\widetilde  T_{\mu\rho}
        \Big\} = 0
\,.
\qquad
\label{energy conservation I}
\end{eqnarray}
Inserting $\widetilde T_{\mu\nu}=\widetilde T^{\scr(0)}_{\mu\nu}+\widetilde T^{\scr(1)}_{\mu\nu}$ from~(\ref{Einstein equation for perturbations})
into this equation yields at the leading order in $\kappa$,
\begin{eqnarray}
\partial^\mu \widetilde T^{\scr(0)}_{\mu\nu} 
\!-\! \mathcal{H}\delta_\nu^0 \eta^{\mu\nu}\widetilde T^{\scr(0)}_{\mu\nu} 
  &\!=\!& 0 \;\Longrightarrow \;
  \partial_\nu\mathcal{L}^{\scr(0)}\!-\!\partial^\mu\mathcal{K}^{\scr(0)}_{\mu\nu}
  \!-\! \mathcal{H}\delta_\nu^0 \big(D\mathcal{L}^{\scr(0)}\!-\!\mathcal{K}^{\scr(0)}\big) = 0
\,,
\qquad
\label{energy conservation: 0th order}
\end{eqnarray}
where $\mathcal{K}^{\scr(0)}=\eta^{\mu\nu}\mathcal{K}^{\scr(0)}_{\mu\nu}$.
Inserting the decomposition~(\ref{Kmn: split}) 
into Eq.~(\ref{energy conservation: 0th order}) one gets,
\begin{equation}
 \partial_\nu\big(\mathcal{L}^{\scr(0)}\!+\!\mathcal{Q}^{\scr(0)}\big)
   \!-\!\delta_\nu^0\partial_0\mathcal{S}^{\scr(0)}
  \!-\! \mathcal{H}\delta_\nu^0 \big[D\big(\mathcal{L}^{\scr(0)}\!+\!\mathcal{Q}^{\scr(0)}\big)\!-\!\mathcal{S}^{\scr(0)}\big] = 0
\,.
\qquad
\label{energy conservation: 0th order 2}
\end{equation}
Upon recalling that $\partial_\nu\rightarrow \delta_\nu^0\partial_0$
and in light of~(\ref{pressure and rho vs L and K}), 
this can be rewritten as, 
\begin{equation}
 \partial_\nu\big(a^D\rho^{\scr(0)}\big)
  \!-\! \mathcal{H}\delta_\nu^0 \big[(D\!-\!1)\mathcal{P}^{\scr(0)}\!-\!\rho^{\scr(0)}\big] = 0
\,,
\qquad
\label{energy conservation: 0th order 3}
\end{equation}
which is equivalent to the leading order conservation law in the standard form, 
\begin{equation}
\partial_0\rho^{\scr(0)}\!+\!(D\!-\!1)\mathcal{H}\big(\rho^{\scr(0)}\!+\!\mathcal{P}^{\scr(0)}\big) 
 = 0
\,.\qquad
\end{equation}
Next we consider the first order equation, which is easily obtained from~(\ref{energy conservation I}),
\begin{eqnarray}
\partial^\mu \widetilde T^{\scr (1)}_{\mu\nu} 
\!-\! \mathcal{H}\delta_\nu^0 \eta^{\mu\nu}\widetilde T^{\scr (1)}_{\mu\nu}
&\!=\!&\kappa\Big[\partial_\mu\big(h^{\mu\rho}\widetilde T^{\scr (0)}_{\rho\nu}\big)
   \!+\!\frac12(\partial_\nu h^{\rho\mu})\widetilde  T^{\scr (0)}_{\rho\mu}
    \!-\!\frac12(\partial^\mu h)\widetilde T^{\scr (0)}_{\mu\nu}
\!-\!\mathcal{H}\delta_\nu^0h^{\mu\rho}\widetilde  T^{\scr (0)}_{\mu\rho}
          \Big]
.
\qquad\;
\label{energy conservation: 1st order}
\end{eqnarray}
Taking account of~(\ref{Einstein equation for perturbations: 1st order}),
according to which,
\begin{eqnarray}
\widetilde{T}_{\mu\nu}^{\scr (1)} 
  &\!=\!& \eta_{\mu\nu}\mathcal{L}^{\scr (1)} \!-\!\mathcal{K}^{\scr (1)}_{\mu\nu}
\nonumber\\
 &\!\!& \hskip -.4cm
  \!+\,\kappa\Big[
                        h_{\mu\nu}\mathcal{L}^{\scr (0)} \!-\! \frac{1}{2}h_{(\mu}^\beta 
\mathcal{K}^{\scr (0)}_{\nu)\beta} 
\!-\! \frac12\eta_{\mu\nu} h^{\alpha\beta} \mathcal{K}^{\scr (0)}_{\alpha\beta}
\!+\!\frac14 \epsilon^{\tau\sigma\beta\delta}\big(\partial_\sigma h_{\beta(\mu}\big)
  \eta_{\nu)\delta}\gamma_\tau \gamma^5 iS^{\scr (0)}_{{\tt bb}}(x;x)
                \Big]
,\qquad\;\;
\label{Tmn d1st order}
\end{eqnarray}
the conservation law~(\ref{energy conservation: 1st order}) can be rewritten as, 
\begin{eqnarray}
\partial_\nu \mathcal{L}^{\scr (1)}
\!-\! \partial^\mu  \mathcal{K}^{\scr (1)}_{\mu\nu}
\!-\! \mathcal{H}\delta_\nu^0\big(D\mathcal{L}^{\scr (1)}
 \!-\!\eta^{\mu\nu}\mathcal{K}^{\scr (1)}_{\mu\nu}\big) 
&\!=\!&-\kappa\bigg[\frac34\partial^\mu\big(h_{\mu}^{\rho}\mathcal{K}^{\scr (0)}_{\rho\nu}\big)
   \!-\!\frac14\partial^\mu\big(h^{\rho}_{\nu}\mathcal{K}^{\scr (0)}_{\rho\mu}\big)
\nonumber\\
&&\hskip -0.9cm
  \!-\,\frac12h^{\mu\rho}\big(\partial_\nu \mathcal{K}^{\scr (0)}_{\mu\rho}\big)
    \!-\!\frac12(\partial^\mu h)\mathcal{K}^{\scr (0)}_{\mu\nu}
\!+\!\delta_\nu^0\frac{D\!-\!1}{2}\mathcal{H}h^{\mu\rho}\mathcal{K}^{\scr (0)}_{\mu\rho}
\nonumber\\
&&\hskip -1.5cm
 \!+\,\frac14\partial^\mu\Big(\big(\partial_\sigma h_{\alpha(\mu}\big)\epsilon^{\rho\sigma\alpha}_{\;\;\;\;\;\;\nu)} \gamma_\rho \gamma^5 iS_{\tt bb}^{\scr(0)}(x;x)\Big)
          \bigg]
.
\qquad\;
\label{energy conservation: 1st order II}
\end{eqnarray}
Remarkably, any dependence on $\mathcal{L}^{\scr (0)}$ has dropped out from the 
right-hand side of this equation.

To solve~(\ref{energy conservation: 1st order II}), 
notice first that the terms in the first and third line can be easily absorbed into 
$\mathcal{K}^{\scr (1)}_{\mu\nu}$. Indeed, upon writing 
\begin{equation}
 \mathcal{K}^{\scr (1)}_{\mu\nu}= \kappa\bigg[\frac34h_{\mu}^{\rho}\mathcal{K}^{\scr (0)}_{\rho\nu}
   \!-\!\frac14h^{\rho}_{\nu}\mathcal{K}^{\scr (0)}_{\rho\mu}
   \!+\!\frac14\big(\partial_\sigma h_{\alpha(\mu}\big)\epsilon^{\rho\sigma\alpha}_{\;\;\;\;\;\;\nu)} \gamma_\rho \gamma^5 iS_{\tt bb}^{\scr(0)}(x;x)
          \bigg] \!+\! k^{\scr (1)}_{\mu\nu}
          \,,
\label{shifting Kmn}
\end{equation}
Eq.~(\ref{energy conservation: 1st order II}) simplifies to, 
\begin{eqnarray}
\partial_\nu \mathcal{L}^{\scr (1)}
\!-\! \partial^\mu  k^{\scr (1)}_{\mu\nu}
\!-\! \mathcal{H}\delta_\nu^0\big(D\mathcal{L}^{\scr (1)}
 \!-\!\eta^{\mu\nu}k^{\scr (1)}_{\mu\nu}\big) 
&\!\!&\hskip -0cm
\nonumber\\
&&\hskip -2.4cm
=-\kappa\Big[\!-\frac12h^{\mu\rho}\big(\partial_\nu \mathcal{K}^{\scr (0)}_{\mu\rho}\big)
    \!-\!\frac12(\partial^\mu h)\mathcal{K}^{\scr (0)}_{\mu\nu}
    \!+\!\delta_\nu^0\frac{D}{2}\mathcal{H}h^{\mu\rho}\mathcal{K}^{\scr (0)}_{\mu\rho}
          \Big]
,
\qquad\;\;
\label{energy conservation: 1st order III}
\end{eqnarray}
where the parity violating term in~(\ref{shifting Kmn}) drops out, as its trace vanishes.
One can make further progress towards solving~(\ref{energy conservation: 1st order II})
by making use of the decomposition~(\ref{Kmn: split}), upon which the right-hand-side 
simplifies to, 
\begin{eqnarray}
(RHS)
=-\kappa\bigg[\frac12\partial_\nu \big(h \mathcal{Q}^{\scr (0)}\big)
    \!+\!\frac12h_{00}\big(\partial_\nu \mathcal{S}^{\scr (0)}\big)
    \!-\!\frac12\delta_\nu^0(\partial_0h)\mathcal{S}^{\scr (0)}
    \!-\!\delta_\nu^0\frac{D}{2}\mathcal{H}\big(h_{00}\mathcal{S}^{\scr (0)}\!+\!h\mathcal{Q}^{\scr (0)}\big)
          \bigg]
,\hskip -0.2cm
\nonumber\\
\label{energy conservation: 1st order IV}
\end{eqnarray}
Now, the first term can be clearly absorbed $\mathcal{L}^{\scr(1)}$. Indeed, writing 
\begin{equation}
 \mathcal{L}^{\scr (1)}= -\frac{\kappa}{2}h\mathcal{Q}^{\scr (0)} \!+\! \ell^{\scr (1)}
          \,,
\label{shifting L}
\end{equation}
Eq.~(\ref{energy conservation: 1st order III}) further simplifies to,
\begin{eqnarray}
\partial_\nu \ell^{\scr (1)}
\!-\! \partial^\mu  k^{\scr (1)}_{\mu\nu}
\!-\! \mathcal{H}\delta_\nu^0\big(D\ell^{\scr (1)}
 \!-\!\eta^{\mu\nu}k^{\scr (1)}_{\mu\nu}\big) 
=-\kappa\delta_\nu^0\bigg[
    \frac12h_{00}\big(\partial_0\mathcal{S}^{\scr (0)}\big)
    \!-\!\frac12(\partial_0h)\mathcal{S}^{\scr (0)}
    \!-\!\frac{D}{2}\mathcal{H}h_{00}\mathcal{S}^{\scr (0)}
          \!\bigg]
,\hskip -0.4cm
\nonumber\\
\label{energy conservation: 1st order V}
\end{eqnarray}
where we took account of the fact that $\mathcal{S}^{\scr (0)}$ does not depend on spatial 
coordinates (due to the translational symmetry of the background).
No local solution in the graviton fields to~(\ref{energy conservation: 1st order V}) exists. 
Indeed, making 
the {\it Ansatz},
\begin{equation}
k^{\scr (1)}_{\mu\nu} = - \delta_\mu^0\delta_\nu^0 s^{\scr (1)}
\,,\qquad 
\ell^{\scr (1)} = 0
\,,\qquad
\label{Ansatz for k1 and l1}
\end{equation}
reduces~(\ref{energy conservation: 1st order V}) to,
\begin{eqnarray}
a\partial_0\frac{s^{\scr (1)}}{a}
=\frac{\kappa}{2}\bigg[
    h_{00}\big(\partial_0\mathcal{S}^{\scr (0)}\big)
    \!-\!(\partial_0h)\mathcal{S}^{\scr (0)}
    \!-\!D\mathcal{H}h_{00}\mathcal{S}^{\scr (0)}
          \!\bigg]
\,,\qquad
\label{energy conservation: 1st order VI}
\end{eqnarray}
whose general solution can be written as, 
\begin{eqnarray}
s^{\scr (1)}(\eta,\vec x)
=\kappa a(\eta) \int_{\eta_0}^\eta\frac{{\rm d}\eta^\prime}{2a(\eta')}
    \big[h_{00}\big(\partial_0\mathcal{S}^{\scr (0)}\big)
    \!-\!(\partial_0h)\mathcal{S}^{\scr (0)}
    \!-\!D\mathcal{H}h_{00}\mathcal{S}^{\scr (0)}
          \big](\eta^\prime,\vec x)
\,,\qquad
\label{energy conservation: 1st order VII}
\end{eqnarray}
which can be added to~(\ref{shifting Kmn}).

To summarize, we have found out that the perturbed energy-momentum 
tensor~(\ref{Tmn d1st order}) acquires through
$\eta_{\mu\nu}\mathcal{L}^{\scr (1)} \!-\!\mathcal{K}^{\scr (1)}_{\mu\nu}$
additional local and non-local contributions which, when incorporated into 
$\widetilde{T}_{\mu\nu}^{\scr (1)}$, results in, 
\begin{eqnarray}
\widetilde{T}_{\mu\nu}^{\scr (1)} 
  &\!=\!&\kappa\Big[ h_{\mu\nu}\mathcal{L}^{\scr (0)}\!-\!\frac12 \eta_{\mu\nu}h\mathcal{Q}^{\scr (0)}
  - h_{\mu}^{\rho}\mathcal{K}^{\scr (0)}_{\nu\rho}
  \!-\! \frac12\eta_{\mu\nu} h^{\alpha\beta} \mathcal{K}^{\scr (0)}_{\alpha\beta}
  \!+\! \delta_\mu^0\delta_\nu^0 s^{\scr (1)}
   \Big]
\,.\qquad\;
\label{Tmn d1st order II}
\end{eqnarray}
which can be also written as, 
\begin{eqnarray}
\widetilde{T}_{\mu\nu}^{\scr (1)} 
  &\!=\!&\kappa\Big[h_{\mu\nu}\big(\mathcal{L}^{\scr (0)}\!+\!\mathcal{Q}^{\scr (0)}\big)
       \!-\!\delta_{(\mu}^0 h_{\nu)0}\mathcal{S}^{\scr (0)}
  \!+\! \frac12\eta_{\mu\nu} h_{00} \mathcal{S}^{\scr (0)}
  \!+\! \delta_\mu^0\delta_\nu^0 s^{\scr (1)}
   \Big]
\,.\qquad\;
\label{Tmn d1st order III}
\end{eqnarray}
Notice that the parity violating term dropped out completely from 
$\widetilde{T}_{\mu\nu}^{\scr (1)}$. More importantly, 
the dynamical graviton contributes in~(\ref{EE: 1st order no derivatives RHS}) as,
\begin{equation} 
  \frac{\kappa^2}{2}a^{2-D} h_{\mu\nu}\big(\mathcal{L}^{\scr (0)}\!+\!\mathcal{Q}^{\scr (0)}\big)
              =\frac{\kappa^2}{2}a^{2} h_{\mu\nu}\mathcal{P}^{\scr (0)}
\,,\qquad\;
\label{Tmn d1st order IV}
\end{equation}
which is precisely what is needed to cancel the would-be dynamical graviton mass term
in~(\ref{EE: 1st order no derivatives RHS}), resolving the dynamical graviton 
mass problem in the linear equation of motion for the gravitational perturbations 
on general cosmological backgrounds in general $D$.

However, when~(\ref{Tmn d1st order IV}) is inserted 
into~(\ref{EE: 1st order no derivatives RHS}), one sees that 
the gravitational potential ($h$ and $h_{00}$) and vector ($h_{0i}$) perturbations do not 
cancel out from the no-derivative terms 
in~(\ref{EE: 1st order no derivatives LHS})--(\ref{EE: 1st order no derivatives RHS}),
so the question of scalar (and vector) perturbations persists.

\medskip
{\bf Einstein tensor conservation.} Note first that 
the term proportional to $\Lambda_0$ in~(\ref{Einstein equation for perturbations})
multiplies $g_{\mu\nu} = a^2(\eta_{\mu\nu} + \kappa h_{\mu\nu})$, from 
which it immediately follows that $\nabla^\mu g_{\mu\nu} = 0 $, and 
therefore that term is covariantly conserved. Next, we have also checked 
by explicit calculation that $\nabla^\mu G_{\mu\nu} = 0$,
both at the zeroth and first order in $\kappa$, 
making a nontrivial check of the correctness of $G_{\mu\nu}^{\scr (0)}$
and $G_{\mu\nu}^{\scr (1)}
=\kappa \mathcal{L}_{\mu\nu}^{\;\;\;\;\rho\sigma}h_{\rho\sigma}$
given in Eqs.~(\ref{Einstein tensor: 0th order}) 
and~(\ref{Lichnerowicz operator}), respectively. 
In other words, the graviton equation of motion~(\ref{Einstein equation for perturbations})
 is (covariantly) transverse, as it should be. 
Since these equations constitute a contracted Bianchi identity, 
which is satisfied for arbitrary metric perturbations, 
they provide no further information for the problem at hand.


\section{Simple model of fermionic backreaction}
\label{Simple model of fermionic backreaction}

As a simple example, in this section we consider the problem of 
the fermion backreaction in the simple case in which fermions are 
in thermal equilibrium, and Universe's expansion can be considered as adiabatic.
This means that the temperature scales with the scale factor as, 
$T\propto 1/[a g^\frac13_{* s}(a)]$, where $g_{* s}(a)$ is the number of relativistic 
degrees of freedom in the plasma, which only changes when some species 
become nonrelativistic. Neglecting this dependence, 
we have $\dot T/T \simeq H\ll  T$, so that $T$ can be considered as 
an adiabatic function of time, and $\dot T$ can be neglected.

With this in mind, we can write the leading order thermal fermionic propagator 
governed by the equation of motion~(\ref{Dirac equation zeroth order})
 as ({\it cf.} Ref.~\cite{BarrosoMancha:2020fay}), 
\begin{equation}
i\widetilde{S}^{\scr(0)}(x;x') = \big(i\gamma^\mu\partial_\mu + a M^\dagger\big)
   i\Delta_F(x;x')
   \,,
\label{Fermionic propagator: 0th order}
\end{equation}
where 
\begin{eqnarray}
   i\Delta_F(x;x')  = \frac{(am)^{D-2}}{(2\pi)^{D/2}}
   \frac{K_\frac{D-2}{2}\big(am\sqrt{\Delta x^2_{++}}\,\big)}
        {\big(am\sqrt{\Delta x^2_{++}}\,\big)^\frac{D-2}{2}}
        -\int \frac{{\rm d}^3 k}{(2\pi)^3}{\rm e}^{i \bm{k}\cdot (\bm{x} - \bm{x'})}
        \frac{\cos[\omega(t\!-\!t')]}{\omega\big({\rm e}^{\beta \omega}\!+\!1\big)}
   \,,\quad\;
\label{Fermionic propagator: 0th order 2}
\end{eqnarray}
where $\omega = \sqrt{\|\bm{k}\|^2+(am)^2}$, 
$\Delta x^2_{++} = -(|\eta\!-\!\eta'|\!-\!i\epsilon)^2 \!+\! \|\bm{x}\! -\! \bm{x'}\|^2$,
and
we set $D = 4$ limit in the thermal part of the propagator,
as it is finite in $D=4$.
This solution neglects the difference between particles and antiparticles 
in Eqs.~(\ref{Ansatz: fermionic 2pt fn: 0th order})--(\ref{equations for particles and antiparticles}), which 
is justified when $H\ll m$, which is what we assume here.

The scalar part of the propagator~(\ref{Fermionic propagator: 0th order 2}) 
 evaluates  at space-time coincidence to, 
\begin{equation}
   i\Delta_F(x;x)  = \frac{(am)^{D-2}}{(4\pi)^{D/2}}
   \Gamma\Big(1\!-\!\frac{D}{2}\Big)
        +\frac{1}{2\pi^2\beta^3(am)}
        \left[\partial_z J_F(4,z)\right]_{z= \beta am}
   \,,\qquad
\label{Fermionic propagator: 0th order 3}
\end{equation}
where $J_F(4,z)$ denotes the fermionic thermal integral,
\begin{equation}
   J_F(n,z) = \int_0^\infty\! {\rm d} x\, x^{n-2} \ln\!\left[
      1\!+\!\exp\big(
      {-\sqrt{x^2\!+\!z^2}}\,\,\big)\right]  
   \,,\qquad
\label{Fermionic propagator: 0th order 4}
\end{equation}
whose argument, $z = am/T$ can be reintepreted as a temperature 
that scales inversely with the scale factor, {\it i.e.} $T(t) = T/a(t)$, where
$T$ denotes the initial temperature (we work in units in which 
the Boltzmann constant, $k_B=1$). 

Next, we evaluate $\mathcal{K}^{\scr (0)}_{\mu\nu}$ defined in 
Eq.~(\ref{definition of Kmn and L: 2PI}),
\begin{eqnarray}
\mathcal{K}_{\mu\nu}^{{\scr(0)}{\tt bb}}(x;x)&=&
-\frac{i}2 {\rm Tr}\Big\{\gamma_\nu \Big[\big(\partial_\mu \!-\! \partial_\mu^\prime\big)
i S^{{\tt bb}}(x';x) \Big]_{x'\rightarrow x}\,\Big\}
\nonumber\\
 &\!=\!& -2^{\frac{D}{2}-1}\Big[\big(\partial_\mu\!-\!\partial_\mu^\prime\big)\partial_\nu
  i\Delta_F(x;x')\Big]_{x'\rightarrow x}
\,,\qquad
\label{evaluation of Kmn}
\end{eqnarray}
where we made use of, ${\rm Tr}\big(\gamma_\nu\gamma^\alpha\big)
 = -{\rm Tr}[\mathbb{I}]\delta_\nu^\alpha 
 = -2^\frac{D}{2}\delta_\nu^\alpha$.
 Making use of,
 \begin{equation}
 \frac{K_{\nu}(z)}{z^\nu} = \frac{\Gamma(-\nu)}{2^{\nu+1}} \sum_{n=0}^\infty
   \frac{(z/2)^{2n}}{(\nu+1)_nn!}
   +\frac{\Gamma(\nu)}{2^{\nu+1}} \sum_{n=0}^\infty
   \frac{(z/2)^{2n-2\nu}}{(-\nu+1)_nn!}
 \,,\quad \Big(\nu = \frac{D\!-\!2}{2}\,\Big)\,,\quad
 \label{evaluation of Kmn 2}
 \end{equation}
and of the fact, that the $D$-dependent series does not contribute at 
and near coincidence, one finds that~(\ref{evaluation of Kmn}) evaluates to, 
\begin{eqnarray}
\mathcal{K}_{\mu\nu}^{{\scr(0)}{\tt bb}}(x;x)&=&
 \frac{(am)^{D}}{2(2\pi)^{D/2}}
   \Gamma\Big(\!\!-\!\frac{D}{2}\Big)\eta_{\mu\nu}
        +\frac{2\eta_{\mu\nu}}{3\pi^2\beta^4}
        \left[\frac{1}{z}\partial_z J_F(4,z)\right]_{z= \beta am}
\nonumber\\
 &&
 +\, \frac{2\delta_{\mu}^0\delta_{\nu}^0}{3\pi^2\beta^4}
  \left[\frac{4}{z}\partial_z J_F(6,z)
  \!+\!3z\partial_z J_F(4,z)\right]_{z= \beta am}
\,,\qquad
\label{evaluation of Kmn 3}
\end{eqnarray}
where, when evaluating the momentum integral, we used 
$k_i k_j\rightarrow \delta_{ij}\|\bb{k}\|^2/3$.
Noting that,
\begin{equation}
\frac{1}{z}\partial_z J_F(n,z) = -(n\!-\!3)J_F(n\!-\!2,z)
\,,\quad (n> 3)
\,,\qquad
\label{derivative of JF}
\end{equation}
Eq.~(\ref{evaluation of Kmn 3}) simplifies to, 
\begin{eqnarray}
\mathcal{K}_{\mu\nu}^{{\scr(0)}{\tt bb}}(x;x)&=&
 \frac{(am)^{D}}{2(2\pi)^{D/2}}
   \Gamma\Big(\!\!-\!\frac{D}{2}\Big)\eta_{\mu\nu}
        -\frac{2\eta_{\mu\nu}}{\pi^2\beta^4}
        \left[ J_F(4,z)\right]_{z= \beta am}
\nonumber\\
 &&
 -\, \frac{2\delta_{\mu}^0\delta_{\nu}^0}{\pi^2\beta^4}
  \left[4 J_F(4,z)
  \!+\!z^2J_F(2,z)\right]_{z= \beta am}
\,.\qquad
\label{evaluation of Kmn 4}
\end{eqnarray}
Inserting this and~(\ref{Fermionic propagator: 0th order 3})
 into~(\ref{definition of Kmn and L: 2PI}) gives,  
\begin{equation}
\mathcal{L}^{{\scr(0)}{\tt bb}}(x;x) 
= \eta^{\mu\nu}\mathcal{K}^{{\scr(0)}{\tt bb}}_{\mu\nu}(x;x)
  + a {\rm Tr}\big[Mi S^{{\scr(0)}{\tt bb}}(x;x) \big]
   = 0
\,,
\label{evaluation of L}
\end{equation}
which was to be expected as $\mathcal{L}^{{\scr(0)}{\tt bb}}(x;x)$
represents the leading order  on-shell contribution to the Lagrangian. 

We are now ready to calculate the energy-momentum tensor~(\ref{Einstein equation for perturbations}). The leading order contribution is,
\begin{eqnarray}
T_{\mu\nu}^{{\scr (0)}{\tt b}}
&\!=\!& a^{2-D}
\Big(\eta_{\mu\nu} \mathcal{L}^{{\scr (0)}{\tt bb}}(x;x)
  \!-\! \mathcal{K}^{{\scr (0)}{\tt bb}}_{\mu\nu}(x;x)
  \Big)
\\
&\!=\!& \hskip -0.cm
\bigg[\!-\! \frac{m^{D}}{2(2\pi)^{D/2}}
   \Gamma\Big(\!\!-\!\frac{D}{2}\Big)
        \!+\!\frac{2}{\pi^2(\beta a)^4}
        \left[ J_F(4,z)\right]_{z= \beta am}
        \bigg](a^2\eta_{\mu\nu})
\nonumber\\
 &&\hskip 0.1cm
 +\, \frac{2a^2\delta_{\mu}^0\delta_{\nu}^0}{\pi^2(\beta a)^4}
  \left[4 J_F(4,z)
  \!+\!z^2J_F(2,z)\right]_{z= \beta am}
\,.
\label{Tmn: leading order}
\end{eqnarray}
The divergence $\propto 1/(D-4)$ in~(\ref{Tmn: leading order}) 
can be renormalized by the cosmological constant 
counterterm action in~(\ref{HE action}), 
which contributes to the energy-momentum tensor as, 
\begin{eqnarray}
T_{\mu\nu}^{\,{\rm ct}}
&\!=\!&
 -\frac{2}{\sqrt{-g^{\scr (0)}}}
        \frac{\delta S_{\rm\scr HE}} {\delta g^{\mu\nu}}
=  T_{\mu\nu}^{\scr (0){\rm ct}} +  T_{\mu\nu}^{\scr (1){\rm ct}} +\mathcal{O}(\kappa^2)
\nonumber\\
&\!=\!&
        -\frac{2}{\kappa^2}\left[G_{\mu\nu}
        \!+\!\frac{D\!-\!2}{2}\Lambda_0g_{\mu\nu}\right]
\,,
\label{Tmn: counterterm}
\end{eqnarray}
where $g_{\mu\nu} = a^2\big(\eta_{\mu\nu}+\kappa h_{\mu\nu}\big)$.
Recalling that,
\begin{equation}
 \Gamma\Big(\!\!-\!\frac{D}{2}\Big) = -\frac{1}{D\!-\!4}
  \left[1\!+\!\frac{D\!-\!4}{2}\Big(\gamma_{\rm\scr E}\!-\!\frac{3}{2}\,\Big)\right]
  \!+\!\mathcal{O}(D\!-\!4)
  \,,
\label{expanding Gamma}
\end{equation}
where $\gamma_{\rm\scr E}= -\psi(1)=0.57\dots$ is the Euler constant,
we see that adding $T_{\mu\nu}^{\scr (0){\rm ct}}$  to (\ref{Tmn: leading order}) 
with a cosmological constant,
$\Delta \Lambda \equiv \Lambda_0 -\Lambda$,
which -- in the minimal subtraction scheme -- amounts to,
\begin{equation}
-\frac{D\!-\!2}{\kappa^2}\Delta\Lambda = -\frac{m^4}{8\pi^2}
\times \frac{\mu^{D-4}}{D\!-\!4}
 \,,\qquad
\label{Lambda0}
\end{equation}
renormalizes the action~(\ref{Tmn: leading order}). The renormalized energy-momentum
tensor is obtained by summing the two contributions, 
\begin{eqnarray}
T_{\mu\nu}^{{\scr (0)}{\rm ren}}&\!=\!&T_{\mu\nu}^{{\scr (0)}{\tt b}}+T_{\mu\nu}^{{\scr (0)}{\rm ct}} 
\nonumber\\
&\!=\!&\left[\frac{m^4}{16\pi^2}\left( \ln\left(\frac{m^2}{2\pi\mu^2}\right) 
         \!+\!\gamma _{\rm\scr E}\!-\!\frac{3}{2}\right)
 \!+\!\frac{2}{\pi^2(\beta a)^4}
        \left[ J_F(4,z)\right]_{z= \beta am}
\right]g^{\scr(0)}_{\mu\nu}
\label{Tmn: leading order 3a}\\
 &+\!&\hskip 0.cm
 \left[\frac{2}{\pi^2(\beta a)^4}
  \left[4 J_F(4,z)
  \!+\!z^2J_F(2,z)\right]_{z= \beta am}\right]
   \big(a^2\delta_{\mu}^0\delta_{\nu}^0\big)
\,.
\label{Tmn: leading order 3b}
\end{eqnarray}
The first three terms in~(\ref{Tmn: leading order 3a}) are the usual one-loop vacuum
contributions to the vacuum energy, and (if one wants) they can be be subtracted by choosing
the finite cosmological constant to be, 
\begin{equation}
\frac{\Lambda}{\kappa^2}= \frac{m^4}{16\pi^2}
\left( \ln\left(\frac{m^2}{2\pi\mu^2}\right) 
         \!+\!\gamma _{\rm\scr E}\!-\!\frac{3}{2}\right)
 \,.\qquad
\label{Lambda fin}
\end{equation}
%
The remaining terms in~(\ref{Tmn: leading order 3a})--(\ref{Tmn: leading order 3b})
are the termal one-loop contributions, and have the form of 
a perfect cosmological fluid~(\ref{perfect fluid}) 
with pressure and energy density given by,
\begin{eqnarray}
\mathcal{P}^{(0)} &\!=\!& \frac{2}{\pi^2(\beta a)^4}
        \left[ J_F(4,z)\right]_{z= \beta am}
        \nonumber\\
\mathcal{P}^{\scr (0)}\!+\!\rho^{(0)}&\!=\!& \frac{2}{\pi^2(\beta a)^4}
  \left[4 J_F(4,z)
  \!+\!z^2J_F(2,z)\right]_{z= \beta am}
\,.\qquad
\label{perfect fluid from fermions}
\end{eqnarray}

Consider now the first order contributions 
in~(\ref{Einstein equation for perturbations: 1st order}).
Neglecting for the moment  
$\mathcal{L}^{\scr (1)}(x;x)$ and $\mathcal{K}^{\scr (1)}_{\mu\nu}(x;x)$,
one obtains,
\begin{eqnarray}
T^{\scr (1)}_{\mu\nu} 
&\!\supset\!& a^{2-D}\kappa
\bigg\{\frac12
\Big[
h_{\mu\nu}\!+\!\eta_{\mu\nu} h
                 \Big]\bigg[
                 \!-\!\frac{(am)^{D}}{2(2\pi)^{D/2}}
   \Gamma\Big(\!\!-\!\frac{D}{2}\Big)
        \!+\!\frac{2}{\pi^2\beta^4}
        \left[ J_F(4,z)\right]_{z= \beta am}
                         \bigg]
\nonumber\\
 &&\hskip 0cm
 \!+\,
\Big[
h_{(\mu}^0 \delta_{\nu)}^0
\!+\!\eta_{\mu\nu} h_{00}
                 \Big]
                                  \bigg[\frac{1}{\pi^2\beta^4}
  \left[4 J_F(4,z)
  \!+\!z^2J_F(2,z)\right]_{z= \beta am}
                 \bigg]
\bigg\}
\,,\quad\,
\label{Tmn: 1st order: main part}
\end{eqnarray}
where we took account of~(\ref{evaluation of Kmn 4}), (\ref{evaluation of L})
and~(\ref{Fermionic propagator: 0th order 3}).
Notice that, in the adiabatic approximation we are working in, the parity violating term 
in~(\ref{Einstein equation for perturbations: 1st order}) dropped out.~\footnote{The parity violating term in~(\ref{Einstein equation for perturbations}) could in principle contribute
a subleading contribution. However, based on 
Eqs.~(\ref{Ansatz: fermionic 2pt fn: 0th order}) and~(\ref{equations for particles and antiparticles}) we see that, upon taking a trace, its contribution vanishes,
\begin{eqnarray}
&&\frac{a^{2-D}}{4}\kappa \epsilon^{\tau\sigma\beta\delta}\big(\partial_\sigma h_{\beta(\mu}\big)
  \eta_{\nu)\delta}{\rm Tr}\Big[\gamma_\tau \gamma^5 iS^{\scr (0)}_{{\tt bb}}(x;x)\Big]
\nonumber\\
&&\hskip 2cm
=\,\frac{a^{2-D}}{4}\kappa  \epsilon^{\tau\sigma\beta\delta}\big(\partial_\sigma h_{\beta(\mu}\big)
  \eta_{\nu)\delta}\times 2^{\frac{D}{2}-1}\delta_\tau^0 \frac{am_Rm_I}{m}
  \sum_\pm(\pm i\mp i) i\Delta_{\scr(\pm)}^{{\tt bb}}(x;x) = 0
\,.\qquad
\nonumber
\end{eqnarray}
Moreover, the analysis in section~\ref{On-shell analysis} suggests that, 
when the first order contributions from
$\mathcal{L}^{\scr (1)}(x;x)$ and $\mathcal{K}^{\scr (1)}_{\mu\nu}(x;x)$
are added, the parity violating term drops out on-shell. 
}

Comparing the order $\kappa$ counterterm in~(\ref{Tmn: counterterm})
 with the structure of the divergences in the leading order~(\ref{Tmn: leading order})
and in the first order term~(\ref{Tmn: 1st order: main part}) show that 
the counterterm~(\ref{Tmn: counterterm}) cannot be used to renormalize both divergences 
(because the coefficient of the divergent contribution in~(\ref{Tmn: 1st order: main part})
 is by a factor 1/2 too small). The expectation is that, including the remaining contributions 
  to the energy-momentum tensor from 
$\mathcal{L}^{\scr (1)}(x;x)$ and $\mathcal{K}^{\scr (1)}_{\mu\nu}(x;x)$,
would render the problem renormalizable.
The analysis of section~\ref{Conservation laws come to the rescue} shows that 
the total first order energy-momentum tensor can be written as~(\ref{Tmn d1st order II}), 
(\ref{Tmn d1st order III}). Inserting ~(\ref{evaluation of Kmn 4}) into~(\ref{Tmn d1st order II})
gives,  
\begin{eqnarray}
T_{\mu\nu}^{\scr (1)} 
  &\!=\!&\kappa \big(a^2 h_{\mu\nu}\big)\bigg[\!-\!\frac{m^{D}}{2(2\pi)^{D/2}}
   \Gamma\Big(\!\!-\!\frac{D}{2}\Big)
     \!+\!\frac{2}{\pi^2(a\beta)^4}\Big[J_F(4,z)\Big]_{z=\beta(am)}
         \,\bigg]
\nonumber\\
&&   \!+\,\kappa a^{2-D}\Big[\!-\!\delta_{(\mu}^0 h_{\nu)0}\mathcal{S}^{\scr (0)}
  \!+\! \frac12\eta_{\mu\nu} h_{00} \mathcal{S}^{\scr (0)}
  \!+\! \delta_\mu^0\delta_\nu^0 s^{\scr (1)}
   \Big]
\,,\qquad\;
\label{Tmn first order calculated}
\end{eqnarray}
where 
\begin{equation}
\mathcal{S}^{\scr(0)} = \frac{2}{\pi^2\beta^4}\Big[4J_F(4,z)\!+\!z^2J_F(2,z)\Big]_{z=\beta(am)}
\,,\qquad
\label{S0: calculated}
\end{equation}
and $ s^{\scr (1)}$ is the thermal contribution in Eq.~(\ref{energy conservation: 1st order VII}).

By comparing~(\ref{Tmn: leading order}) with~(\ref{Tmn first order calculated})
one sees that the cosmological constant counterterm~(\ref{Tmn: counterterm}), with the choice~(\ref{Lambda0}), renormalizes 
the energy-momentum tensor to the linear order in $\kappa$. The renormalized 
energy-momentum tensor can be written as, 
\begin{eqnarray}
T_{\mu\nu}^{{\rm ren}}&\!=\!&T_{\mu\nu}^{{\scr (0)}{\tt b}}
\!+\!T_{\mu\nu}^{{\scr (1)}{\tt b}}
\!+\!T_{\mu\nu}^{{\rm ct}} 
\nonumber\\
&\!=\!&\left[\frac{m^4}{16\pi^2}\left( \ln\left(\frac{m^2}{2\pi\mu^2}\right) 
         \!+\!\gamma _{\rm\scr E}\!-\!\frac32\right)
 \!+\!\frac{2}{\pi^2(\beta a)^4}
        \left[ J_F(4,z)\right]_{z= \beta am}
\right]g_{\mu\nu}
\nonumber\\
 &&\hskip 0.cm
+\,\kappa\left[\frac{2}{\pi^2(\beta a)^4}
  \left[4 J_F(4,z)
  \!+\!z^2J_F(2,z)\right]_{z= \beta am}\right]
   \big(a^2\delta_{\mu}^0\delta_{\nu}^0\big)
\nonumber\\
&& +\,\kappa  a^2 \bigg[\!\!-\!\delta_{(\mu}^0 h_{\nu)0}
  \!+\! \frac12\eta_{\mu\nu} h_{00}
   \bigg]
   \frac{2}{\pi^2(\beta a)^4}\Big[4J_F(4,z)\!+\!z^2J_F(2,z)\Big]_{z=\beta(am)}
 \nonumber\\
&&  +\,\frac{\kappa}{a^2}\delta_\mu^0\delta_\nu^0 s^{\scr (1)}
.\qquad\;\;
\label{renormalized Tmn}
\end{eqnarray}
This can be then expanded in the high and low temperature limits.~\footnote{Useful high temperature expansions ($z\ll 1$)  are~\cite{Quiros:2007zz},
\begin{equation}
J_F(4,z) \simeq \frac{7\pi^4}{360}-\frac{\pi^2}{24}z^2
-\frac{z^4}{32}\ln\Big(\frac{z^2}{a_F}\Big)
\,,\quad 
4J_F(4,z)+z^2J_F(2,z) \simeq\frac{7\pi^4}{90}-\frac{\pi^2}{12}z^2+\frac{z^4}{16}
\,,\quad
\nonumber
\end{equation}
with $a_F = \pi^2\exp(\tfrac{3}{2}-\gamma_E)$, $\ln(a_F) \simeq2.6351$. 
In the low temperature limit ($z\gg 1$) we have, 
\begin{equation}
J_F(4,z) \simeq \sqrt{\frac{\pi z^3}{2}} {\rm e}^{-z}
\,,\quad 
4J_F(4,z)+z^2J_F(2,z) \simeq\sqrt{\frac{\pi z^3}{2}} \big(4+z\big){\rm e}^{-z}
\,.\quad
\nonumber
\end{equation}
}
We have thus shown that, while the na\^{i}ve energy-momentum tensor was not 
renormalizable, 
adding consistently the first order corrections derived in section~\ref{Conservation laws come to the rescue} renders the energy-momentum tensor renormalisable.

The renormalized energy-momentum tensor~(\ref{renormalized Tmn}) 
is an improvement when compared with the naive one in~(\ref{Tmn: 1st order: main part}) 
as it allowed for consistent renormalization. However, it is fair to say that we still lack a rigorous framework within which one can study the evolution of 
gravitational perturbations through different epochs of the early Universe.


\section{Conclusion and Outlook}
\label{Conclusion and Outlook}

In this work we have studied Einstein's gravity~(\ref{HE action}) coupled to 
massive Dirac fermions (\ref{Dirac action}), where -- to mimic the standard model -- we included both
scalar and pseudoscalar masses~(\ref{fermionic mass}). Our goal is to understand the dynamics of gravitational perturbations in general cosmological backgrounds driven by fermionic 
matter.~\footnote{With some effort, our analysis can be adapted to other types of matter present in 
the standard model,
such as scalar and vector fields~\cite{Fennema:2024}.}
The first step is to expand the gravitational and fermionic actions to the second order in gravitational perturbations around a general (spatially flat) cosmological background~(\ref{metric: 0th order}), 
resulting in Eqs.~(\ref{HE: 0th order})--(\ref{HE: 2nd order}) 
and~(\ref{Dirac action: 0th order})--(\ref{Dirac action: 2nd order}).
The 2PI form of the fermionic action, which is more convenient for applications in cosmology,
is given in~(\ref{Dirac action: 0th order: 2PI})--(\ref{Dirac action: 1 loop: 2PI}).
The second order action in gravitational perturbations contains non-derivative terms~(\ref{Dirac action: 2nd order: 2PI b})--(\ref{HE plus Dirac : 2nd order: Keldysh}) that do not 
vanish on-shell, {\it i.e.} when the background equations of motion are used, suggesting that 
there is an on-shell graviton mass. While this observation is unexpected, that does not 
mean that there is a physical graviton mass.

In order to clarify this question, in section~\ref{On-shell analysis} we study
the equation of motion for linear gravitational perturbations~(\ref{Einstein equation for perturbations}).
The fermionic energy-momentum tensor contains both zeroth order terms in the gravitational fields,
which are local, and first order terms, which are nonlocal. Including only the local contributions
does not solve the question of the graviton mass, implying that the non-local contributions are important.
Moreover, from the analysis of section~\ref{On-shell analysis} it is unclear how to consistently identify
the local contribution that would cancel the graviton mass.

A better understanding of that question is advanced in 
section~\ref{Conservation laws come to the rescue}, in which we use conservation of 
the energy-momentum tensor to show that 
an additional local contribution to the energy-momentum tensor emerges, which 
removes the dynamical graviton mass. Moreover, while the na\^{i}ve (perturbative) energy-momentum 
is non-renormalizable, the improved 
one~(\ref{Tmn d1st order II})--(\ref{Tmn d1st order III}) is renormalizable, as is shown in 
section~\ref{Simple model of fermionic backreaction}.

\begin{figure}[h]
\centering
\vskip -0.4cm
\includegraphics[width=5cm]{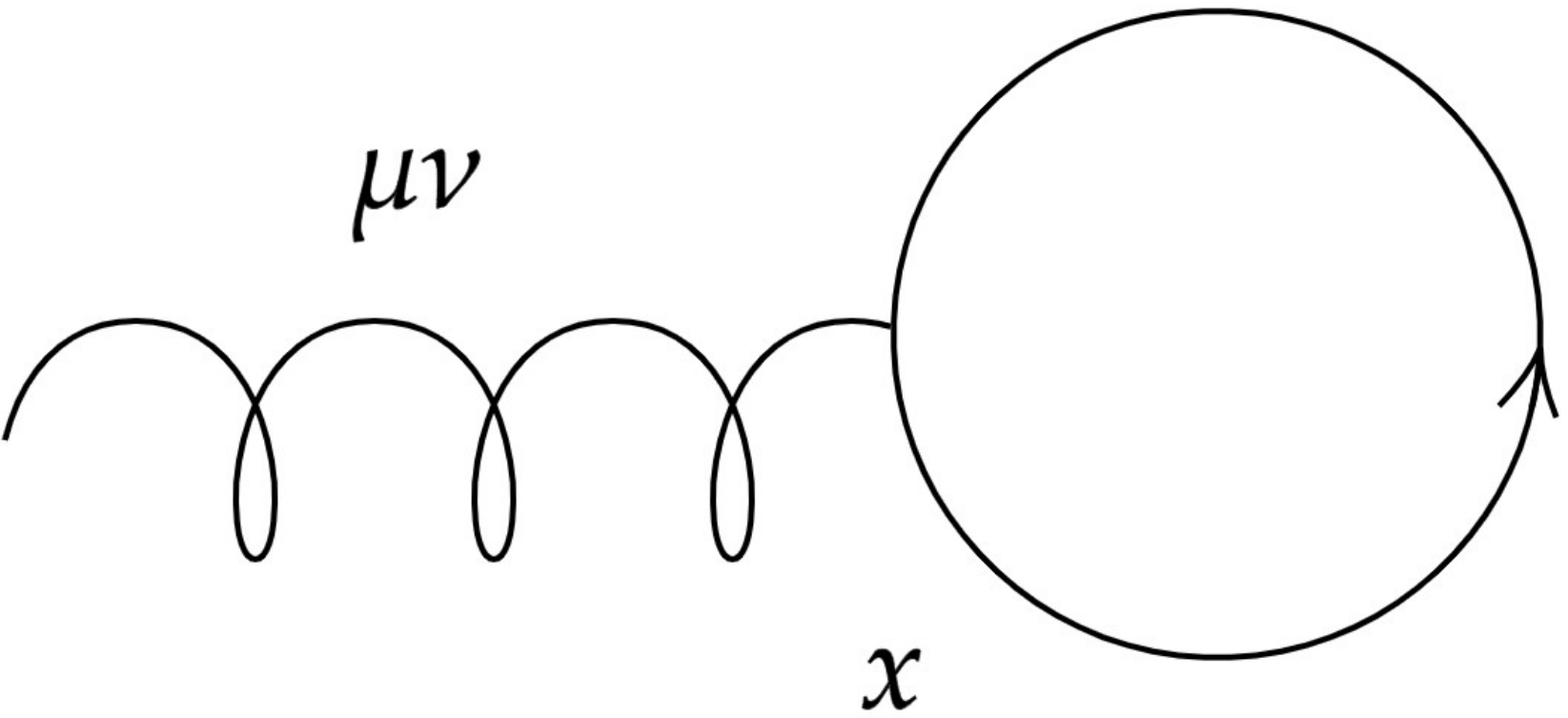}
\vskip -0.7cm
\caption{\footnotesize The one-loop graviton tadpole diagram contributing to the background equation of motion.}
\label{Diagram tadpole}
\end{figure}

While we now do understand how to solve some of the problems that arise when 
studying the dynamics of linear gravitational perturbations on cosmological backgrounds,
a general framework is still lacking, and that is what we discuss next. A systematic framework
for studying the dynamics of gravity on curved backgrounds governed by the action,
\begin{equation}
S[g_{\mu\nu},\psi] = S_{\rm\scr g}[g_{\mu\nu}] +S_{\rm\scr m}[g_{\mu\nu},\psi] 
\,,
\label{total action}
\end{equation}
where $S_{\rm\scr g}[g_{\mu\nu}]$ denotes the gravitational action, 
{\it e.g.}  the Hilbert-Einstein action~(\ref{HE action}), and $S_{\rm\scr m}[g_{\mu\nu},\psi]$ is the matter action ($\psi$ denotes any matter field), {\it e.g.} the Dirac action~(\ref{Dirac action}),
can be obtained in the context of perturbative QFT as follows.
Firstly, one expands the gravitational field $g_{\mu\nu}$ into a background field $g^{\scr(0)}_{\mu\nu}$
and perturbations $\delta g_{\mu\nu}$:  
$g_{\mu\nu}=g^{\scr(0)}_{\mu\nu}+\delta g_{\mu\nu}$. The corresponding expansion
for cosmological backgrounds is ({\it cf.} Eqs.~(\ref{Weyl transformation})--(\ref{tilde gmn})),
\begin{equation}
g_{\mu\nu}(x) =  a^2\left(\eta_{\mu\nu} +\kappa h_{\mu\nu}\right)
\,,\qquad \big( g^{\scr(0)}_{\mu\nu} = a^2\eta_{\mu\nu},\;
 \delta g_{\mu\nu} = a^2\kappa h_{\mu\nu}\big)
\,.\quad\;
\end{equation}
The desired set of dynamical equations consists of two equations, which can be obtained 
by applying the background field method and a perturbative expansion 
to the action~(\ref{total action}). The first equation is the dynamical background equation,  and it is in the form of
the semi-classical Einstein equation~\cite{Birrell:1982ix},
\begin{eqnarray}
G_{\mu\nu}^{\scr(0)}\!+\!\Lambda g_{\mu\nu}^{\scr(0)} 
&\!=\!& 8\pi G \big[T_{\mu\nu}^{\rm\scr cl} 
     +  \big\langle \mathbb{T}^*\big[\hat T_{\mu\nu}\big]\big\rangle
     \big]_{g_{\mu\nu}=g^{\small(0)}_{\mu\nu}}
     \,,\qquad
\label{semiclassical gravity}
\end{eqnarray}
where $G_{\mu\nu}^{\scr(0)} =\left[ \frac{\kappa^2}{\sqrt{-g}}\frac{\delta S_{\rm\scr g}}{\delta g_{\mu\nu}}\right]_{g_{\mu\nu}=g^{\scr(0)}_{\mu\nu}}$ is the Einstein tensor for the background metric $g_{\mu\nu}^{\scr(0)}$,
$T_{\mu\nu}^{\scr cl} $ and $\big\langle \mathbb{T}^*\big[\hat T_{\mu\nu}\big]\big\rangle$
are the classical and quantum contributions to the energy-momentum tensor, 
and $ \mathbb{T}^*$ stands for $ \mathbb{T}$-star time ordering, according to which vertex derivatives (if any) are pulled out of the time-ordered two-point function contributing to the energy-momentum tensor. Eq.~(\ref{semiclassical gravity}) makes physical sense only when
$\big\langle \mathbb{T}^*\big[\hat T_{\mu\nu}\big]\big\rangle$ is renormalized,
{\it  i.e.} all the divergences it contains are removed by local counterterms, 
and it is well known how to do that in semiclassical gravity~\cite{Birrell:1982ix}.
$\big\langle \mathbb{T}^*\big[\hat T_{\mu\nu}\big]\big\rangle$ can be evaluated 
in some perturbative scheme, {\it e.g.} the one-loop truncation, for which 
the corresponding diagram is shown in figure~\ref{Diagram tadpole}.
While it is not known how to solve~(\ref{semiclassical gravity}) self-consistently
on general gravitational backgrounds, 
different approximation schemes have been advanced
by Anderson~\cite{Anderson:2002hh,Anderson:2011wq} and others~\cite{Tsamis:2005je,Anderson:2015yda,Anderson:2020hgg,Pla:2020tpq,Glavan:2013mra,Glavan:2014uga,Glavan:2015cut}, but only the technique of stochastic inflation adapted to late time cosmology~\cite{Glavan:2017jye,Belgacem:2021ieb,Belgacem:2022fui,Vedder:2022spt} 
allows for a consistent solution of semiclassical gravity~(\ref{semiclassical gravity}) in cosmological settings.

\begin{figure}[h]
\centering
\includegraphics[width=15cm]{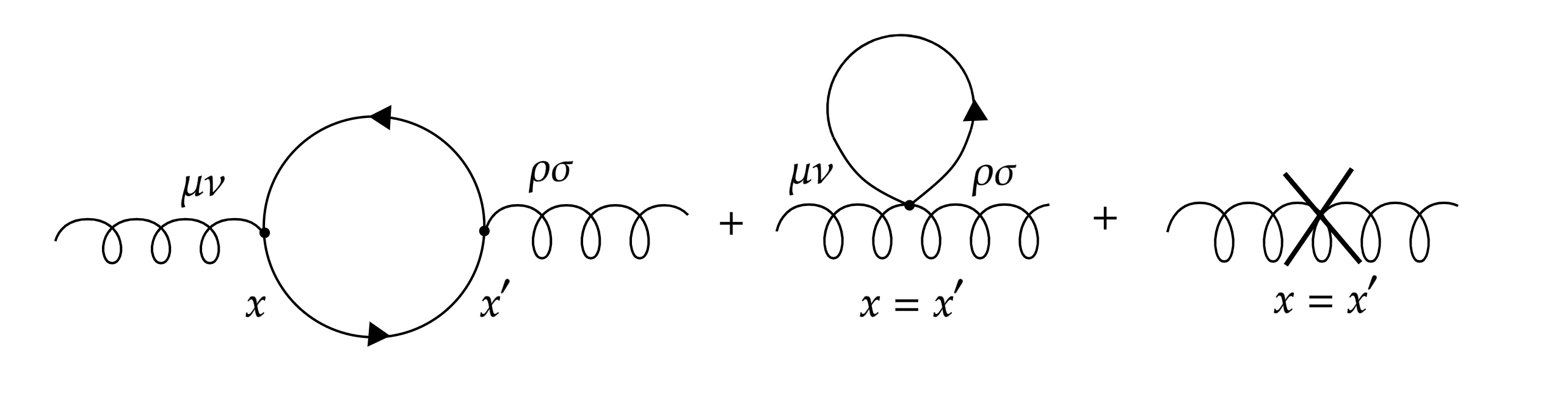}
\vskip -0.3cm
\caption{\footnotesize The one-loop diagrams contributing to the graviton self-energy 
induced by fermions. The last diagram represents the local counterterms
obtained from the counterterm action~(\ref{counterterm action}).}
\label{Diagram selfenergy}
\end{figure}

The second equation governs the evolution of gravitational perturbations, and it can be written as,
\begin{eqnarray}
&&\hskip -0.3cm
\mathcal{L}_{\rm\scr cov}^{\mu\nu\rho\sigma}(x)\delta g^{{\tt\scr b}}_{\rho\sigma}(x) 
  \!+\!\sum_{\tt\scr c}\!{\tt c}\!\!\int\!\! {\rm d}^4x^\prime\sqrt{\!-g^{\scr(0)}(x^\prime)}
   \,\big[{}^{\mu\nu}_{\tt\scr b}\Sigma_{{\tt\scr c}}^{\rho\sigma}\big](x;x^\prime)
                      \delta g^{{\tt\scr c}}_{\rho\sigma}(x^\prime) 
\!+\!\mathcal{O}\big((\delta g^{{\tt\scr c}}_{\rho\sigma})^2\big)  = 0
\,,\qquad\;
\label{perturbation equation: selfenergy}
\end{eqnarray}
where $\mathcal{L}_{\rm\scr cov}^{\mu\nu\rho\sigma}(x)$ denotes the (covariant) 
Lichnerowicz operator~(\ref{Lichnerowicz operator}), 
\begin{equation}
\mathcal{L}_{\rm\scr cov}^{\mu\nu\rho\sigma}(x)\times\frac{\delta^4(x-x^\prime)}
 {\sqrt{-g}}
 \equiv \left[\frac{\kappa^2}{\sqrt{-g(x)}\sqrt{-g(x^\prime)}} 
 \frac{\delta^2 S_{\scr g}}{\delta g_{\mu\nu}(x) \delta g_{\rho\sigma}(x')}\right]_{g_{\mu\nu}=g^{\scr(0)}_{\mu\nu}}
 \,,\quad
\label{Lichnerowicz operator: cov}
\end{equation}
and $\big[{}^{\mu\nu}_{\;\;\,{\tt\scr b}}\Sigma_{{\tt\scr c}}^{\rho\sigma}\big](x;x^\prime)$ is the graviton
self-energy which is, in the Schwinger-Keldysh formalism and in the one-loop truncation, given by,
\begin{eqnarray}
&& \hskip -.2cm
\big[{}_{\mu\nu}^{\;\;\,{\tt\scr b}}\Sigma^{{\tt\scr c}}_{\rho\sigma}\big](x;x^\prime)
= \frac{\kappa^2}{\sqrt{-g_{{\tt\scr b}}}\sqrt{-g^\prime_{{\tt\scr c}}}} 
\bigg\langle\!\! 
\left. \bigg\{ i \frac{\delta S_{\scr m}}{\delta g_{{\tt\scr b}}^{\mu\nu}(x)} \frac{\delta S_{\scr m}}{\delta g_{{\tt\scr c}}^{\rho\sigma} (x')} 
\!+\! \frac{\delta^2 S_{\scr m}}{\delta g_{{\tt\scr b}}^{\mu\nu}(x) \delta g_{{\tt\scr c}}^{\rho\sigma}(x')} \bigg\} \right\rvert_{g_{\mu\nu}=g^{\scr(0)}_{\mu\nu}}\! \bigg\rangle
, \qquad \;
\label{eq:gginter2}
\end{eqnarray}
where $g_{{\tt\scr b}}=g_{{\tt\scr b}}(x)$ and 
$g^\prime_{{\tt\scr c}}=g_{{\tt\scr c}}(x^\prime)$.
This equation has been much less studied than~(\ref{semiclassical gravity})
 in cosmological settings~\cite{Tsamis:1996qk,Park:2010pj,Park:2011ww,Leonard:2014zua,Park:2015kua,Tan:2021ibs,Miao:2024nsz,Miao:2024atw,Kavanagh:2024,Fennema:2024}, 
 but it is as important for our purpose.
Eq.~(\ref{perturbation equation: selfenergy}) is written in the Schwinger-Keldysh notation,
so that it can be used both for evolution of the graviton one-point function (in which case 
the Keldysh indices on $ \delta g^{{\tt\scr c}}_{\rho\sigma}$ can be dropped) and 
for evolution of the graviton two-point functions, in which case 
one ought to multiply~(\ref{perturbation equation: selfenergy}) from the right by 
$\delta g^{{\tt\scr d}}_{\rho\sigma}(x^{\prime\prime})$ and take an expectation value~\footnote{Alternatively, Eq.~(\ref{perturbation equation: selfenergy: 2pt fn}) 
can be obtained by varying the suitably truncated 2PI effective action 
which includes the graviton one- and two-point functions.} 
to obtain a set of evolution equations for the graviton two-point functions
$i\left[{}_{\mu\nu}^{\;\;\,{\tt\scr b}}\Delta^{{\tt\scr c}}_{\gamma\delta}\right](x;x^\prime)$,
\begin{eqnarray}
&&\hskip -0.3cm
\mathcal{L}_{\rm\scr cov}^{\mu\nu\rho\sigma}(x)
\,i\!\left[{}_{\rho\sigma}^{\;\;\,{\tt\scr b}}\Delta^{{\tt\scr c}}_{\gamma\delta}\right](x;x^\prime) 
\nonumber\\
&&\hskip 0.6cm
  \!+\sum_{\tt\scr c}\!{\tt c}\!\int\! {\rm d}^4x^{\prime\prime}
  \sqrt{\!-g^{\scr(0)}(x^{\prime\prime})}
   \,\big[{}^{\mu\nu}_{\;\;\,{\tt\scr b}}\Sigma_{{\tt\scr c}}^{\rho\sigma}\big](x;x^{\prime\prime})
                      \,i\!\left[{}_{\rho\sigma}^{\;\;\,{\tt\scr c}}\Delta^{{\tt\scr d}}_{\gamma\delta}\right](x^{\prime\prime};x^\prime) 
= i\hbar {\tt c}\delta^{\tt\scr cd}\delta^4(x\!-\!x^\prime)
\,,\qquad\;
\label{perturbation equation: selfenergy: 2pt fn}
\end{eqnarray}
where $i\!\left[{}_{\mu\nu}^{\;\;\,{\tt\scr b}}\Delta^{{\tt\scr c}}_{\gamma\delta}\right]\!(x;x^\prime)$ ($\tt b,c = \pm$) are the graviton two-point functions. For example, 
when $\tt b = \mp$ and  $\tt c = \pm$ one obtains the positive and
negative frequency graviton Wightman functions,
\begin{equation}
i\!\left[{}_{\mu\nu}^{\;\;\,{\tt\scr -}}\Delta^{{\tt\scr +}}_{\gamma\delta}\right]\!(x;x^\prime)
   =  \left\langle\delta \hat{g}_{\mu\nu}(x) 
   \delta \hat{g}_{\gamma\delta}(x^\prime) \right\rangle
\,,\quad\;
i\!\left[{}_{\mu\nu}^{\;\;\,{\tt\scr +}}\Delta^{{\tt\scr -}}_{\gamma\delta}\right]\!(x;x^\prime)
   =  \left\langle\delta \hat{g}_{\gamma\delta}(x^\prime) 
   \delta \hat{g}_{\mu\nu}(x) \right\rangle
\,.\quad\;
\label{graviton 2pt functions}
\end{equation}
Just as in the case of the background equation~(\ref{semiclassical gravity}), 
the self-energy in Eq.~(\ref{perturbation equation: selfenergy}) must be renormalized.
Renormalization of the one-loop graviton self-energy demands at least four 
geometric counterterms~\cite{Kavanagh:2024},
\begin{equation}
S_{\rm ct}[g_{\mu\nu}] = \int {\rm d}^D x \sqrt{-g} \left(c_1 R^2 
 \!+\!c_2 W_{\mu\nu\rho\sigma}W^{\mu\nu\rho\sigma}
  \!+\!c_3 R \!+\!c_4  \right)
  \,,\qquad 
\label{counterterm action}
\end{equation}
where $W_{\mu\nu\rho\sigma}$ denotes the Weyl curvature tensor,~\footnote{One can equivalently use the Riemann tensor counterterm action, 
$\int {\rm d}^D x \sqrt{-g} c^\prime_2 R_{\mu\nu\rho\sigma}R^{\mu\nu\rho\sigma}$.} with the goal to remove all nontransverse parts
by local counterterms. The corresponding one-loop Feynman
diagrams for the graviton self-energy are shown in figure~\ref{Diagram selfenergy}.
\vskip -0.2cm
Gravity is a gauge theory, so both equations~(\ref{semiclassical gravity})
and~(\ref{perturbation equation: selfenergy}) 
(or equivalently~(\ref{perturbation equation: selfenergy: 2pt fn})) 
must in addition satisfy certain
consistency conditions. Thus Eq.~(\ref{semiclassical gravity}) 
must obey the following  transversality conditions, 
\begin{equation}
\nabla_{\!\scr(0)}^\mu\, G^{\scr(0)}_{\mu\nu}=0
\,,\quad\;
\nabla_{\!\scr(0)}^\mu\, g^{\scr(0)}_{\mu\nu}=0
\,,\quad\;
\nabla_{\!\scr(0)}^\mu\, T_{\mu\nu}^{\rm\scr cl} =0
\,,\quad\;
\nabla_{\!\scr(0)}^\mu  \big\langle \mathbb{T}^*
\big[\hat T_{\mu\nu}\big]\big\rangle
     \big]_{g_{\mu\nu}=g^{\!\scr(0)}_{\mu\nu}}=0
\,,
\label{Bianchi and conservation}
\end{equation}
which, in the presence of matter field condensates, may require a 
delicate analysis~\cite{Glavan:2023lvw}. The first condition 
in~(\ref{Bianchi and conservation}) is the contracted Bianchi identity, and must hold for
an arbitrary metric, and the last two are the energy-momentum conservation laws, 
and the classical and quantum contributions must be separately conserved.
Likewise, there are two transversality conditions that are expected to hold 
for Eq.~(\ref{perturbation equation: selfenergy}),
\begin{equation}
\nabla^{\scr(0)}_\mu \mathcal{L}_{\rm\scr cov}^{\mu\nu\rho\sigma}(x)=0
\,,\quad\;
\nabla^{\scr(0)}_\mu i\big[{}^{\mu\nu}_{\;\;\,{\tt\scr b}}\Sigma_{{\tt\scr c}}^{\rho\sigma}\big](x;x^\prime) =0
\,,\quad\;
\label{transversality and Ward identity}
\end{equation}
The former identity follows immediately from the contracted 
Bianchi identity in~(\ref{Bianchi and conservation}) 
and the definition of
the Lichnerowicz operator~(\ref{Lichnerowicz operator: cov}), while the
latter is the Ward identity for the graviton self-energy~\cite{Capper:1973pv,Capper:1974vb,Grillo:1999yw,Burns:2014bva}.
Notice that the graviton two-point functions~(\ref{graviton 2pt functions})
need not be transverse (they are transverse only in the special class of
exact de Donder gauges~\cite{Mora:2012zi}).

Experience shows~\cite{Kavanagh:2024,Fennema:2024}
that a transverse self-energy is obtained only after tuning the cosmological constant 
counterterm such that the one-loop matter fluctuations 
do not contribute an additional cosmological constant. This observation is particularly important when considering the question of the graviton mass, as it tells us that 
the graviton masslessness is protected by the gauge symmetry it satisfies.
While at the quantum level transversality of the graviton self-energy 
is imposed by the Ward (or Slavnov-Taylor) identity, at the classical level it is the second Noether identity
that guarantees transversality of the evolution equation for gravitational perturbations.
The classical equation is obtained as the classical limit of 
the quantum equation~(\ref{perturbation equation: selfenergy})  
for the one-point function, which can be obtained from~(\ref{perturbation equation: selfenergy}) by setting 
$\delta g^{{\tt\scr b}}_{\mu\nu}(x) \rightarrow \delta g_{\mu\nu}(x)$,
\begin{eqnarray}
&&\hskip -0.3cm
\mathcal{L}_{\rm\scr cov}^{\mu\nu\rho\sigma}(x)\delta g_{\rho\sigma}(x) 
  \!+\!\int\!\! {\rm d}^4x^\prime\sqrt{\!-g^{\scr(0)}(x^\prime)}
   \,\big[{}^{\mu\nu}\Sigma_{\rm ret}^{\rho\sigma}\big](x;x^\prime)
                      \delta g_{\rho\sigma}(x^\prime) 
\!+\!\mathcal{O}\big((\delta g_{\rho\sigma})^2\big)  = 0
\,,\qquad\;
\label{perturbation equation: selfenergy: class}
\end{eqnarray}
where 
\begin{equation}
\big[{}^{\mu\nu}\Sigma_{\rm ret}^{\rho\sigma}\big](x;x^\prime) 
= \big[{}^{\mu\nu}_{\;\;\,{\tt\scr +}}\Sigma_{{\tt\scr +}}^{\rho\sigma}\big](x;x^\prime)
\!-\!\big[{}^{\mu\nu}_{\;\;\,{\tt\scr +}}\Sigma_{{\tt\scr -}}^{\rho\sigma}\big](x;x^\prime)
\qquad
\label{retarded self-energy}
\end{equation}
is the retarded self-energy. There are however physical situations in which 
one can dynamically generate vacuum energy. Examples of such processes in 
the early Universe setting are the electroweak transition, in which the mass generation
mechanism also changes the vacuum energy, and the strong transition,
induced by chiral quark condensates which also change the vacuum energy.
This means that one can choose to tune the vacuum energy to zero 
either before or after such a transition, but not both, implying that 
the self-energy cannot be made transverse in both regimes. 
The question what are the ramifications of this fact for the dynamics of gravitational
perturbations will be addressed elsewhere~\cite{Fennema:2024}.

The principal message of this work is that 
one cannot obtain consistent evolution equations for (classical or quantum) 
gravitational perturbations on cosmological backgrounds generated by 
quantized matter fields without properly quantizing matter both 
in the semiclassical equation for gravity~(\ref{semiclassical gravity}),
but also in the equation of motion for perturbations on quantized matter 
backgrounds~(\ref{perturbation equation: selfenergy}). These equations are studied in more detail in the example of scalar electrodynamics in Ref.~\cite{Fennema:2024}.

\section*{Acknowledgments}

The authors thank Rick Vinke, whose master thesis was
used as the point of departure and inspiration for this work.
This work is part of the Delta ITP consortium,
a program of the Netherlands Organisation for Scientific Research (NWO) that is
funded by the Dutch Ministry of Education, Culture and Science (OCW) — NWO
project number 24.001.027.
LL is funded by NSFC grant NO. 12165009, Hunan Natural Science Foundation NO. 2023JJ30487.


\end{document}